\documentclass[11pt]{article}

\usepackage{amssymb,amstext,amsmath,amsthm}
\usepackage[dvips]{graphicx}
\usepackage{latexsym}
\usepackage{psfrag}
\usepackage{amsfonts}
\usepackage{bbm}
\usepackage{color}

\newcommand{\Ref}[1]{(\ref{#1})}

\def\nn{\nonumber}

\newcommand{\eqa}{\begin{eqnarray}}
\newcommand{\neqa}{\end{eqnarray}}
\newcommand{\equ}{\begin{equation}}
\newcommand{\nequ}{\end{equation}}

\def\om{\omega}
\def\w{\wedge}

\newcommand{\p}{\partial}
\newcommand{\n}{{D}}

\def\Pp{P_{\scr (+)}}\def\Pm{P_{\scr (-)}}

\def\d{\delta}
\def\f{\frac}
\def\tl{\tilde}

\usepackage{bbm}

\newcommand{\scr}{\rm\scriptscriptstyle}

\newcommand{\su}[1]{\mathfrak{su({#1})}}
\newcommand{\so}[1]{\mathfrak{so({#1})}}

\let\eps=\epsilon

\def\Si{\Sigma}

\newcommand{\GN}{G_{\scr N}}
\newcommand{\al}{\alpha}
\newcommand{\be}{\beta}\newcommand{\ga}{\gamma}\newcommand{\La}{\Lambda}\newcommand{\la}{\lambda}

\newcommand{\BsB}{\langle B \w \star B\rangle}
\def\rd{{\rm d}}

\begin{document}

\title{\Large\bf Bi-metric theory of gravity from the non-chiral Plebanski action}
\author{
Simone Speziale
\medskip \\
{\small 
\emph{Centre de Physique Th\'eorique,\footnote{Unit\'e Mixte de Recherche (UMR 6207) du CNRS et des Universites Aix-Marseille I, Aix-Marseille II et du Sud Toulon-Var. Laboratoire affili\'e \`a la FRUMAM (FR 2291).} CNRS-Luminy Case 907, 13288 Marseille Cedex 09, France}}}

\date{\today}

\maketitle

\begin{abstract}
We study a modification of the Plebanski action for general relativity, which leads to a modified theory of gravity with eight degrees of freedom. We show how the action can be recasted as a bi-metric theory of gravity, and expanding around a bi-flat background we identify the six extra degrees of freedom with a second, massive graviton and a scalar mode.
\end{abstract}

\section{Introduction}

It is an intriguing fact that general relativity can be formulated using polynomial actions, cubic in the fundamental fields. One such action, introduced by Plebanski \cite{Plebanski} and revisited in \cite{Capo2}, uses as fundamental fields a connection in the algebra $\mathfrak{g}$ of the local gauge group of gravity, and an antisymmetric tensor, or 2-form, from which a metric is singled out through the imposition of suitable constraints via a Lagrange multiplier.
The action exists in two different versions, the original, chiral one \cite{Plebanski,Capo2} where $\mathfrak{g}=\su{2}$ is the self-dual part of the Lorentz algebra,\footnote{This formulation is often called general relativity in self-dual variables.} and a non-chiral one \cite{Capo2,Mike, De Pietri} where $\mathfrak{g}=\so{3,1}$. These actions are related to the Ashtekar variables \cite{Ashtekar} for loop quantum gravity, and taken as a starting point in the construction of spin foam models \cite{Perez}.

An interesting aspect of these actions is that they admit a natural modification in which the cosmological constant is turned into a function, $\La(\phi)$, of the Lagrange multiplier $\phi$ present in the theory \cite{Capo3,BengtssonMod,BengtssonP,Krasnov,Krasnov2,KrasnovEff,Lee}.
Any non-constant $\La(\phi)$ gives a different modified theory of gravity, whose physical relevance needs to be investigated. Due to the different nature of the fundamental fields, this modification is a priori unrelated to the more familiar extension of the Einstein-Hilbert action with higher curvature invariants of the metric. Furthermore, the two formulations of the Plebanski action turn out to behave very differently under this modification.
For instance, in the self-dual case there are only two propagating degrees of freedom \cite{Krasnov2} for any choice of $\La(\phi)$, like in general relativity, and unlike actions with higher curvature invariants \cite{Stelle,Deruelle}. 
On the other hand, the modified non-chiral action has extra propagating degrees of freedom. The canonical analysis of \cite{Alexandrov} showed the presence of eight degrees of freedom,\footnote{At least for $\La(\phi)$ with non-degenerate Hessian, see below.} although their physical interpretation was not attempted.
Unravelling the reason why the same type of modification on the two originally equivalent actions has different consequences is the first motivation for this paper. The second is to identify these extra degrees of freedom.

As we hope to clarify during the course of this paper, the origin of the difference is the fact that the non-chiral action is naturally a theory of two metrics, unlike the self-dual one. The modification removes a constraint whose role is to single out a unique metric, hence leaving them both independent and dynamical. 
In particular, we will show how the modified action can be recasted as a bi-metric theory of gravity plus auxiliary scalar fields. The presence of the latter makes the theory different from bigravities studied in the literature \cite{Damour}. 
As a consequence, the physical viability of the theory is an open issue to investigate.
As a first step in that direction, we study a perturbative expansion around the bi-flat background. This allows us
to identify eight degrees of freedom, corresponding to a massless and a massive graviton, plus a scalar field. The situation is thus similar to generic bigravities, and the reason for it is that the auxiliary scalars satisfy algebraic equations and can be integrated out.

The study of the modified self-dual theory has already been under some development \cite{KrasnovEff,Krasnov1,Krasnov3,Krasnov4,Freidel,Ishibashi}. Here we focus on the non-chiral one \cite{Lee,PlebImm}, which has received less attention. Our results follow rather simply from the results in the self-dual theory obtained by Krasnov \cite{KrasnovEff,Krasnov4} and Freidel \cite{Freidel}, nonetheless we try to give a self-contained presentation, in the hope of making the paper accessible also to non-experts on the Plebanski formalism.

The fundamental variables of the Plebanski formalism are a connection $\om$ in $\mathfrak{g}$ and a $\mathfrak{g}$-valued set of 2-forms $B$. The key to the formalism is the fact that the 2-forms $B$ can always be used to introduce a metric, regardless of any constraints. The role of the constraints is rather  
to single out this metric, namely to freeze the remaining components of the 2-forms which do not enter the definition of the metric. 
To be more specific, given an $\su{2}$-valued 2-form $B^i_{\mu\nu}$, $i=1,2,3$, a metric can be defined through the well-known Urbantke formula \cite{Urbantke,Capo2},
\equ\label{gU}
\sqrt{g^{\scr U}} \, g^{\scr U}_{\mu\nu} = \f1{12} \eps_{ijk} \, \eps^{\al\be\ga\d} 
B^{i}_{\mu\al}B^{j}_{\be\ga}B^{k}_{\d\nu}.
\nequ
Notice in this formula the completely antysimmetric tensor $\eps_{ijk}$, the unique singlet in the tensor product of three adjoint representations of SU(2). The $B$ field needs to be complex for this metric to have Lorentzian signature, while a real field yields Euclidean signature. The same mechanism can be applied to both the original Plebanski actions for general relativity and the modified theories. In the original case, the action gives quadratic field equations, some of which are the ``metricity constraints'' which freeze the remaining components of $B^i$ not captured by \Ref{gU}, and the rest reduce to the Einstein equations for \Ref{gU}. If we consider a modification where the constraints are removed, not only we get new field equations, but the extra components are not frozen anymore. The surprise is that this can be done without introducing new degrees of freedom \cite{Krasnov2}, which in particular means that the extra components do not become dynamical. The new field equations do however change the dynamics for the metric, leading to a modified theory of gravity.

In the non-chiral formulation \cite{Capo2,Mike,De Pietri} the fundamental 2-form $B^{IJ}_{\mu\nu}$ is $\so{3,1}$- or $\so{4}$-valued, respectively for Lorentzian and Euclidean signature, and real in both cases. We can straighforwardly generalize \Ref{gU} to this case, but this time \emph{two} possible metrics can be defined. This is simply a consequence of the fact that the tensor product of three adjoint representations of the algebra admits two singlets.
A basis in this two dimensional vector space is provided by the tensors 
$\d_{N[I} \d_{J]MKL}$ and $\d_{N[I} \eps_{J]MKL}$, where $\eps_{IJKL}$
is the completely antisymmetric tensor and we defined the identity $\d_{IJKL} = \f12 (\d_{IK} \d_{JL}-\d_{IL} \d_{JK})$. Accordingly, we have a right-handed Urbantke metric $g^{\scr U(+)}_{\mu\nu}$ and a left-handed $g^{\scr U(-)}_{\mu\nu}$,
\equ\label{gpm}
\sqrt{g^{\scr U(\pm)}} \, g^{\scr U(\pm)}_{\mu\nu} = 
\f1{12} \d_{IN}\left(\d_{JMKL} \pm \f12 \eps_{JMKL}\right) \eps^{\al\be\ga\d} B^{IJ}_{\mu\al}B^{KL}_{\be\ga}B^{MN}_{\d\nu}
\nequ
in the $\so{4}$ case, and similarly in the $\so{3,1}$ case. 
As in the self-dual case, the non-chiral action gives quadratic field equations, some of which are the ``simplicity constraints'' identifying these two metrics with one another and freezing the remaining components of $B^{IJ}$, and the rest reduce to Einstein's equations for the unique metric emerging from the constraints. Removing the constraints as in the modification of \cite{Lee}, the two Urbantke metrics become independent and dynamical. This is at the roots of the bi-metric interpretation.

\medskip

The paper is organized as follows. In the next Section, we review the Plebanski formalism. This review is brief, but has the double ambition of introducing the formalism to non-expert, and to present the simplicity constraints under a perspective that might be new also to experts. This perspective makes it manifest that their role is singling out a unique metric, and will be instrumental to understand the modified theory. The action for the modified Plebanski theory is introduced in Section \ref{SecMod}. In Section \ref{SecBi} we show how it can be reformulated as a bigravity theory for the two metrics \Ref{gpm}, plus auxiliary scalar fields. 
In Section \ref{SecPert} we study the perturbative expansion around the bi-flat solution and identify the local degrees of freedom.
Conclusions and some open questions are collected in the final Section \ref{SecConcl}.

\medskip

Throughout the paper, we take units $16\pi \GN=1$, and use greek letters for spacetime indices and latin letters for internal indices.
For simplicity, we work with Euclidean signature, but formulas can be easily modified to Lorentzian signature. We will comment at places where having a Lorentzian signature has non-trivial consequences.
$[a,b]$ means normalized antisymmetrization, and our normalization of 2-forms is $F=\f12 F_{\mu\nu} {\rm d}x^\mu\w {\rm d}x^\nu$.
We introduce the tensor $\d^{IJ}_{KL} = \f12 (\d^I_K \d^J_L-\d^I_L \d^J_K)$
and the shortand notation $\BsB=\f12 \eps_{IJKL} B^{IJ}\w B^{KL}$.
We use $\star=\f12\eps^{IJ}{}_{KL}$ to indicate the Hodge dual in the internal space. To avoid confusion between this Hodge dual and the spacetime one $\f1{2e}\eps^{\mu\nu}{}_{\rho\sigma}$ in spacetime, we will refer to the positive and negative eigenvectors of the algebra Hodge dual respectively as right- and left-handed components, and those of the spacetime Hodge dual respectively as self-dual and antiself-dual.

\section{Review of the Plebanski mechanism}

Consider the following action, 
\equ\label{S}
S(B,\om,\phi) = \int B^{IJ} \w F_{IJ}(\om) - \f12 \left(\phi_{IJKL} +\f\La6  \eps_{IJKL} \right) B^{IJ} \w B^{KL},
\nequ
where $F(\om)$ is the curvature of an $\mathfrak{so(4)}$ connection $\om$, and $B^{IJ}$ a 2-form with values in the algebra. This action can be called non-chiral, to distinguish it from the original Plebanski action, which is identical but uses $\mathfrak{g}=\su{2}$.
$\La$ is the cosmological constant, and the field $\phi^{IJKL}$ a Lagrange multiplier, symmetric under exchange of the first and the second pair, and antysimmetric within each pair. 
In addition, we impose on $\phi$ the constraint $\eps_{IJKL}\phi^{IJKL}=0.$ 
It has therefore the same symmetries of the Riemann tensor, and 20 components. 
Its variation gives the following equations, known as simplicity constraints,
\equ\label{simpl}
B^{IJ}\w B^{KL} = \f1{12} \eps^{IJKL} \, \BsB.
\nequ
Solutions satisfying the non-degeneracy condition $\BsB \neq 0$ can be divided in two sectors \cite{Mike, De Pietri}, 
\equ\label{Bsols} 
B^{IJ}=\pm (1/2)\eps^{IJ}_{KL}  e^K\w e^L, \qquad B^{IJ}=\pm \, e^I\w e^J.
\nequ 
We will review the derivation below. In both cases, the 20 constraints \Ref{simpl} reduce the initial 36 independent components of $B^{IJ}_{\mu\nu}$ down to 16, parametrized by a tetrad $e^I_\mu$. The first sector corresponds to general relativity, while the second one to a topological theory with no local degrees of freedom. This can be seen looking at the field equations, or more sinthetiquely inserting one of the two solutions back in the action \Ref{S}. 

Inserting $B^{IJ}=(1/2)\eps^{IJ}_{KL} e^K\w e^L$, one obtains
\equ\label{SEC}
S(e,\om) = \int \f12 \eps_{IJKL} e^I\w e^J \w F^{KL}(\om) -2 \,\La \,  e.  \\
\nequ
This is the Einstein-Cartan action, whose equivalence to general relativity is established -- for non-degenerate tetrads -- taking the variation by $\om$. The result is Cartan's structure equation d$_{\om}e=0$, solved by the spin connection $\om^{IJ}_\mu(e)=e^I_\nu \nabla_\mu e^{\nu J}$, which further identifies its curvature $F(\om)$ with the Riemann tensor, through Cartan's second structure equation
\equ\label{C2}
R_{ \mu\nu\rho\sigma}(e) \equiv e_{I\rho} e_{J\sigma} F^{IJ}_{\mu\nu}(\om(e)). 
\nequ
Using these results, \Ref{SEC} gives
\equ
S(e) = \int e ( R - 2 \Lambda).
\nequ
More details can be obtained looking at the field equations (e.g. \cite{Mike,PlebImm}), in particular one finds that on solutions the Lagrange multiplier equals the Weyl tensor. 

For completeness, let us also recall how the second sector leads to a topological theory. Inserting the solution $B^{IJ}=e^I\w e^J$, one obtains
\equ\label{topterm}
S(e,\om) = \int e^I\w e^J \w F_{IJ}(\om) -2 \,\La \,  e.  \\
\nequ
For $\La=0$ we can look again at the $e\neq 0$ sector, and integrating over the connection as above, the first term gives $\eps^{\mu\nu\rho\sigma}R_{\mu\nu\rho\sigma}$, which vanishes thanks to the first Bianchi identities. Hence the field equations vanish identically, thus giving no local degrees of freedom. For more details on the theory defined by \Ref{topterm}, see \cite{Liu}. Finally, the case $\La\neq 0$ only admits solutions in the degenerate sector $e=0$.

\subsection{Self-duality and metricity}
The brief overlook above highlights the role of the simplicity constraints \Ref{simpl} in extracting a tetrad from the $B$ field, for which the general relativity dynamics is then recovered. The derivation can be found in details in the literature \cite{Mike,De Pietri}, but we wish to review it here with a slightly different perspective, which will prove instumental to understand the modified theory.
In particular, we would like to stress that the constraints are not needed to introduce a metric. A metric -- as a matter of fact, two -- can always be introduced through the Urbantke formulas \Ref{gpm}. The role of the constraints is rather to single out a unique metric out of the 36 initial components of $B^{IJ}$. 

To see this, we begin from the isomorphism 
$
\mathfrak{so(4)}\cong \mathfrak{su(2)}\oplus \mathfrak{su(2)}, 
$
which allows one to split the components of an $\mathfrak{so(4)}$ algebra element, like $B^{IJ}$, into right- and left-handed components:
\equ\label{Bdec}
B^{IJ} 
= P_{\scr (+)}^{IJ}{}_i \, B^{i}_{\scr +} + P^{IJ}_{\scr (-)}{}_i \, B^{i}_{\scr -}.
\nequ
Here we denoted $\scr (+)$ and ${\scr (-)}$ respectively the right- and left-handed parts, and we introduced the projectors
\equ\label{Pi}
P_{\eps}^{IJ}{}_i = \eps\d^{IJ}_{0i} + \f12 \eps^{IJ}_{0i}, \qquad \eps=\pm.
\nequ
Now consider a single triple of 2-forms $B^i$, and construct the Urbantke metric \Ref{gU}.
The well-known result by Urbantke \cite{Urbantke} states that if the 3x3 matrix $B^i\w B^j$
is invertible, then  
\equ
\f1{2 \sqrt{g^{\scr U}}}\eps^{\mu\nu}{}_{\rho\sigma} B^i_\eps{}_{\mu\nu} = \eps B^i_\eps{}_{\rho\sigma}.
\nequ
Here the sign $\eps$ depends on the sign of the determinant of $B^i\w B^j$.
This means that the triple $B^i$ is self-dual (or antiself-) with respect to the metric \Ref{gU} -- or any other metric in the same conformal class, since the spacetime Hodge dual $\eps^{\mu\nu}{}_{\rho\sigma}/2\sqrt{g^{\scr U}}$ is invariant under conformal transformations.   
As we now discuss, it is then possible to write explicitly $B^i$ in terms of this metric.

Consider a tetrad $e^I_\mu$ associated with a given metric, and the following Plebanski 2-form \cite{Plebanski, Capo2, Mike}
\equ\label{Sigma}
\Sigma^i_\eps(e) = \eps e^0 \w e^i + \f12 \eps^i{}_{jk} e^j\w e^k.
\nequ
These 2-forms are said to be \emph{metric}, and are self-dual in the spacetime metric $e^I_\mu$, as well as right-handed in the $\mathfrak{so(4)}$ algebra (or antiself-dual and left-handed, for $\eps=-1$). Notice that they only depend on 13 of the 16 components of the tetrad, since the internal direction $e^0$ has been fixed. 
Furthermore, they satisfy
\equ\label{SS}
\Sigma^i_\eps(e) \w \Sigma_\eps^j(e) = \eps \, 2 e \d^{ij},
\nequ
where $e = \det \, e_\mu^I$, and here and in the following factors of $d^4x$ are tacitly assumed.
Since for $e\neq 0$ \Ref{Sigma} form a basis in the space of right- or left-handed 2-forms, we can always decompose $B^i_\eps= c^i{}_j \Si^j_\eps(e)$ as a linear combination of the $\Si^i_\eps$ with arbitrary coefficients $c^i{}_j$ \cite{Krasnov, Capo2, Freidel}. This decomposition is clearly defined up to an SO(3) rotation, as well as a global rescaling $c^i{}_j\mapsto \Omega^{-2}c^i{}_j$, $e_\mu^I\mapsto \Omega e_\mu^I$, thus preserving the total number of independent components in $B^i$, that is $9+13-3-1=18$.

The $B^i$ decomposed in this way are self-dual with respect to $e_\mu^I$, and it is straighforward to show that this metric falls in the same conformal class as the Urbantke metric, since the latter evaluates to 
$g^{\scr U}_\eps{}_{\mu\nu}=(\det c) \, e_\mu^I e_\nu^J \d_{IJ}$.
It is then possible, and indeed convenient, to choose the rescaling freedom $\Omega$ to fix $\det c=\pm1$, so that the Urbantke metric coincides with the metric associated to $e^I_\mu$. In conclusion, one can parametrize
\equ\label{BSigma}
B^i_\eps = \eta b^i_a \Sigma^a_\eps(e),
\nequ
where $\eta=\pm1$ is a sign and the coefficients $b^i_a$ form a unimodular matrix. 
The details of the proofs can be found in \cite{Freidel}, whose notation we follow. 
Both types of indices, $i,j$ and $a,b$, are in $\su{2}$. The different notation is useful to treat the scalars $b^i_a$ as a ``triad'', and to keep track of its inverse, given by
\equ\label{binv}
\hat b^a_i = \f12 \eps^{abc} \eps_{ijk} b^j_b b^k_c.
\nequ
In particular, using \Ref{SS} we have
\equ\label{defm}
B^i_\eps \w B^j_\eps = \eps \, 2 e m^{ij}, \qquad m^{ij} = b^i_a b^j_b \d^{ab},
\nequ
that is the unimodular triad $b^i_a$ is given by the normalized eigenvectors of $m^{ij}$.\footnote{Its 8 components can then be identified with two SO(3) rotation and a two-parameters rescaling.}

The parametrization \Ref{BSigma} plays a key role in investigations of the modified self-dual theory \cite{Krasnov,Ishibashi}.
The idea here is to apply it to the $\mathfrak{so(4)}$ case, as it can be done straighforwardly using the decomposition \Ref{Bdec}.
As the right- and left-handed parts in \Ref{Bdec} are independent, we need independent triads and tetrads, say 
$b^i_a, \bar b^i_a$ and $e^I_\mu, \bar e^I_\mu$. Correspondingly, we take 
\equ
B^i_{\scr +} = b^i_a \Sigma^a_{\scr +}(e), \qquad B^i_{\scr -} = \eta \bar b^i_a \Sigma_{\scr -}^a(\bar e),
\nequ
where we already dropped one sign which proves irrelevant in the following.
To shorten our notation, we use from now on 
$\Sigma(e)\equiv \Sigma_{\scr +}(e)$ and $\bar \Sigma(\bar e)\equiv \Sigma_{\scr -}(\bar e)$.
We then write
\equ\label{ParamGen}
B^{IJ} =  P^{IJ}_{\scr (+)}{}_i \, b^i{}_a \Sigma^a(e) + \eta P^{IJ}_{\scr (-)}{}_i \, \bar b^i{}_a \bar \Sigma^a(\bar e),
\nequ
where the Plebanski 2-forms $\Sigma(e)$ and $\bar \Sigma(\bar e)$ encode the two metrics
$g_{\mu\nu}=e^I_\mu e^J_\nu \d_{IJ}$ and $\bar g_{\mu\nu}= \bar e^I_\mu \bar  e^J_\nu \d_{IJ}$.

The decomposition \Ref{ParamGen} parametrizes $B^{IJ}$ in such a way that the right- and left-handed components of $B^{IJ}$ are also self- and antiself-dual, but \emph{with respect to two independent metrics $g_{\mu\nu}$ and $\bar g_{\mu\nu}$}.
These are precisely the two Urbantke metrics defined in \Ref{gpm}: an explicit calculation gives
\equ\label{gUpmeval}
g^{\scr U(+)}_{\mu\nu} = g_{\mu\nu}, \qquad 
g^{\scr U(-)}_{\mu\nu} = \eta \bar g_{\mu\nu}. 
\nequ

\subsection{Simplicity constraints}\label{SecSC}

We now want to take advantage of the parametrization \Ref{ParamGen} in the Plebanski action \Ref{S}. Thanks to the orthogonality of the two $\mathfrak{su(2)}$ algebras, one has
$B^{IJ} \w F_{IJ}(\om^{IJ}) = B^i \w F_i(\om^i) + \bar B^i \w F_i(\bar \om^i)$,
where $\om^i$ and $\bar \om^i$ are the right- and left-handed parts of the $\mathfrak{so(4)}$ connection, and $F$ their curvature, i.e. $F_i(\om) = \f12 \eps_{ijk} (\rd \om^{jk} + \om^j{}_l\w \om^{lk})$.
Concerning the Lagrange multiplier
$
\phi \in {\bf (2,0)}\oplus{\bf (0,2)}\oplus{\bf (1,1)}\oplus{\bf (0,0)}
$
recall that this can be decomposed into its irreducible representations using the projectors \Ref{Pi},
\eqa
\phi^{IJKL} &=& \varphi^{ij} \Pp^{IJ}{}_i \Pp^{KL}{}{}_j + \bar \varphi^{ij} \Pm^{IJ}{}_i \Pm^{KL}{}_j
+ \nn \\ && + \psi^{ij} \left(\Pp^{IJ}{}_i \Pm^{KL}{}_j+\Pm^{IJ}{}_j \Pp^{KL}{}_i \right)
+ \nn \\ && + \varphi_0 \d^{ij}  \left(\Pp^{IJ}{}_i \Pp^{KL}{}_j+\Pm^{IJ}{}_i \Pm^{KL}{}_j \right).
\neqa
Here $\varphi^{ij}$ and $\bar \varphi^{ij}$ are symmetric and traceless (5 components each), $\psi$ is generic (9 components), and $\varphi_0$ is the scalar part. In terms of these quantities, \Ref{S} reads 
\eqa\label{S2pre}
S(B,\bar B, \om,\bar \om, \varphi,\bar \varphi,\psi,\varphi_0) &=& \int B^i\w F_i(\om) + \bar B^i \w F_i(\bar \om) 
\\ &&- \f12 \bigg[
\varphi_{ij} B^i \w B^j + \bar \varphi_{ij}\bar  B^i \w \bar B^j + \psi_{ij} \left( B^i \w \bar  B^j + \bar B^i \w B^j \right) \nn \\ &&
+ \varphi_0 \d_{ij} \left( B^i \w B^j + \bar B^i \w \bar B^j \right)
+\f\La6\d_{ij} \left( B^i \w B^j - \bar B^i \w \bar B^j \right) \bigg],\nn
\neqa
where we used the orthogonality properties of the projectors, see Appendix \ref{AppN}.

Before studying the constraints, let us use the parametrization \Ref{ParamGen}. 
We denote $m^{ij}$ and $\bar m^{ij}$ the matrices \Ref{defm} for $b^i_a$ and $\bar b^i_a$, and their traces $m\equiv\d_{ij}m^{ij}$ and $\bar m\equiv\d_{ij}\bar m^{ij}$. Then, using \Ref{SS}
we have
\eqa\label{S2}
S(\eta, b,\bar b,e,\bar e, \om,\bar \om, \varphi,\bar \varphi,\psi,\varphi_0) &=& 
\int b^i_a \Si^a\w F_i(\om) + \eta \bar b^i_a \bar \Si^a \w F_i(\bar \om)  \nn \\ && -  
e m^{ij} \varphi_{ij} + \bar e \bar m^{ij} \bar \varphi_{ij}  -
2 \eta \ell^{ij} \psi_{ij} + \nn \\ &&
- \phi_0 \d_{ij} \left( em-\bar e \bar m \right)
-\f1{6}\La\d_{ij} \left( em+\bar e \bar m \right)
\neqa
where we introduced the shorthand notation
\equ
\ell^{ij} \equiv \f18 b^i_a \bar b^j_b \, \eps^{\mu\nu\rho\sigma} \Si^a_{\mu\nu}(e) \bar \Si^b_{\rho\sigma}(\bar e).
\nequ

When we vary by the irreducible components of $\phi^{IJKL}$, we find 20 constraints, which we group according to the irreps ${\bf (2,0)}$, ${\bf (0,2)}$, ${\bf (1,1)}$ and ${\bf (0,0)}$ of $\mathfrak{so(4)}$, obtaining respectively
\begin{align}\label{simpl1}
& m^{ij} = \f13\, m \, \d^{ij}, 
& \bar m^{ij} = \f13 \, \bar m \, \d^{ij},
&& \ell^{ij} = 0, 
&& e \, m = \bar e \, \bar m.
\end{align}
Recalling the unimodularity of $m^{ij}$ and $\bar m^{ij}$, the right- and left-handed equations imply that
$m^{ij} = \bar m^{ij} = \d^{ij}$. Next, the ${\bf (1,1)}$ equations $\ell^{ij} = 0$ mean that the two triples $\Sigma^i(e)$ and $\bar \Sigma^i(\bar e)$ are orthogonal to each other, which implies that the two metrics $e_\mu^I$ and $\bar e_\mu^I$ coincide up to a conformal factor. The latter is fixed to 1 by the final equation, since $m = \bar m=3$. Finally, the sign $\eta$ is remains free.
Overall, the solution of the constraints for non-degenerate metrics is 
$$
\bar g_{\mu\nu} = g_{\mu\nu}, \qquad m^{ij} = \bar m^{ij} = \d^{ij}, \qquad \eta=\pm1.
$$ 
The two sectors $\eta=\pm1$ are simply the topological and gravitational sectors described by \Ref{gUpmsol}.
Plugging the solutions back into \Ref{ParamGen}, one immediately identifies $\eta=1$ with the topological sector and $\eta=-1$ with the gravitational one,
\begin{subequations}\label{gUpmsol}\begin{align}
& B^{IJ}=e^K\w e^L &\mapsto && g^{\scr U(+)}_{\mu\nu} = g^{\scr U(-)}_{\mu\nu} = e^I_\mu e^J_\nu \d_{IJ}, \\
& B^{IJ}=\f12\eps^{IJ}_{KL} e^K\w e^L &\mapsto && g^{\scr U(+)}_{\mu\nu} = -g^{\scr U(-)}_{\mu\nu} = e^I_\mu e^J_\nu \d_{IJ}.
\end{align}\end{subequations}
Hence we also see that the two Urbantke metrics \Ref{gpm} coincide on the topological solution, and are opposite on the physical one.

Summarizing, the decomposition \Ref{ParamGen} shows that prior to imposing the simplicity constraints \Ref{simpl}, the field $B^{IJ}$ can be parametrized in terms of the two Urbantke metrics $g^{\scr U(+)}_{\mu\nu} = g_{\mu\nu}$ and $g^{\scr U(-)}_{\mu\nu} = \eta \bar g_{\mu\nu}$, plus the two auxiliary fields $b^i_a$ and $\bar b^i_a$. 
By rewriting the simplicity constraints in the form \Ref{simpl1} we see that ten of them (the ${\bf(2,0)}\oplus {\bf (0,2)}$ part) freeze the auxiliary fields, and ten of them (the ${\bf(1,1)}\oplus {\bf (0,0)}$ part) identify the two initial metrics with one another.
As we show in the rest of the paper, in the modified theory the constraints on $B^{IJ}$ are removed, and the two metrics become independent dynamical fields. 

Let us conclude this brief overview of the Plebanski formalism with some comments.
\begin{itemize}

\item \emph{Relation to BF theory} 

The original Plebanski action \Ref{S} is of the type ``BF plus constraints''. An interesting aspect of this construction is the fact that BF theory is a topological field theory, without any local degrees of freedom. This is due to the invariance of the action under the shift symmetry 
\equ\label{shift}
B\mapsto B + {\rm d}_\om \eta, 
\nequ
where $\eta$ is a 1-form with values in the algebra. This symmetry (which includes the diffeomorphisms) together with the gauge symmetry also present, guarantee that all solutions can be locally mapped to the trivial one \cite{Blau,Baez}. When the constraints are added, the shift symmetry is broken down to diffeomorphisms, and local degrees of freedom are allowed. Although it might appear counter-intuitive at first that adding constraints one increases the number of degrees of freedom, the reason for this is the fact that in BF theory the $B$ field is just a Lagrange multiplier, imposing the flatness condition $F=0$. Therefore, if conditions are placed on a Lagrange multiplier, the resulting theory is less constrained. Specifically, non-trivial curvature is now allowed.

\item \emph{Self-dual case}

As mentioned above, the original Plebabski formulation uses the same action but $\mathfrak{g}=\su{2}$. This case can be recovered from our analysis above, setting $\bar B^i = \varphi_0 = 0$ in \Ref{S2pre}. In that case one parametrizes the fundamental $B^i$ field as in \Ref{BSigma}, the Lagrange multiplier has only components in the ${\bf (2,0)}$ irrep, and the metricity constraints give $m^{ij}=\d^{ij}$.
Notice that the non-chiral action is \emph{not} the sum of two different actions, respectively purely right- and left-handed: the terms in $\psi$ and $\varphi_0$ mix the two, and are crucial to impose the proportionality of the two metrics, which would otherwise be independent.

\item \emph{Lorentzian signature}

All the formulas can be adapted to Lorentzian signature and $\mathfrak{g}=\so{3,1}$, but there is a caveat: the right- and left-handed projectors are now \emph{complex}, $i \eps \d^{IJ}_{0i}+\f12\eps^{IJ}_{0i}$. As a consequence, the decomposition \Ref{ParamGen} requires the $b$ and $\Si$ fields to be complex as well. 
This does not pose any obstruction (at least at the classical level), but one then needs to add extra conditions for the Lorentzian metrics $g_{\mu\nu}$ and $\bar g_{\mu\nu}$ to be real. Such \emph{reality conditions} were studied in \cite{Capo2} for the original action, and revisited in the work by Krasnov (e.g. \cite{KrasnovEff}) for the modified theory. They can be imposed independently on the right- and left-handed sectors. The simplicity constraints then guarantee that also $B^{IJ}$ is real.\footnote{A word of criticism is probably due here. The original rationale for introducing the non-chiral Plebanski action was precisely to eliminate the additional reality conditions  needed in the self-dual action. Our idea of using the algebra decomposition to understand the modified non-chiral theory has the advantage of leading to a simple interpretation as a bigravity theory, as we will show below. However, it forces us to reintroduce the reality conditions to deal with Lorentzian signature. Although this can be seen as a drawback of our approach, the fact that reality conditions can be ultimately dealt with \cite{KrasnovEff} seems not to compromise it.}

\end{itemize}

\section{Modified Plebanski theory}\label{SecMod}

The modification of gravity we consider is obtained promoting the cosmological constant  $\La$ in \Ref{S} to a potential $\La(\phi)$, 
\label{SMod}\equ\label{Smod1}
S(B,\om,\phi) = \int B^{IJ} \w F_{IJ}(\om) - \f12 \left(\phi_{IJKL} +\f16\La(\phi)\eps_{IJKL} \right) B^{IJ} \w B^{KL}.
\nequ
An action of this type has been first introduced in \cite{Lee}, and it extends to the non-chiral action \Ref{S} the modification proposed by Krasnov for the self-dual Plebanski theory \cite{Krasnov}, in turn related to previous work by Capovialle \cite{Capo3} and by Bengtsson and Peldan \cite{BengtssonMod,BengtssonP}.
The field equations obtained varying $\phi$, $B$ and $\om$ are
\begin{subequations}\label{FE}\begin{align}\label{FE1}
& {\rm d}_\om B^{IJ} = \rd B^{IJ} + [\om, B]^{IJ}= 0, \\
& B^{IJ}\w B^{KL} = \left(\f1{12} \eps^{IJKL} - \f13 \f{\d \La(\phi)}{\d \phi_{IJKL}} \right) \, \BsB, \label{FE2} \\
& F^{IJ}(\om) = \phi^{IJKL} B_{KL} + \f16 \La(\phi) \eps^{IJKL} B_{KL}.\label{FE3}
\end{align}\end{subequations}
The first equation is the same compatibility condition of $\om$ with $B$, unchanged from the original Plebanski action. The last two differ from Plebanski's when $\La(\phi)$ is not a constant, thus leading to departures from general relativity. Their nature, and the scale at which they occur, depend upon the specific form of $\La(\phi)$.\footnote{For the reader unfamiliar with the Plebanski formalism, we recall that for $\La$ constant, after the solution $B = \star e \w e$ to the simplicity constraints \Ref{newFE2} is chosen, \Ref{newFE3} splits into equations identifying $\phi$ as the Weyl part of the Riemann tensor, and equations giving the Einstein dynamics to the metric $e^I_\mu$. See \cite{Mike,PlebImm,KrasnovPleb} for more details.}

The key difference with the original action lies in the modified simplicity constraints \Ref{FE2}: since the Lagrange multiplier appears explicitly, these \emph{are not anymore constraints on} $B^{IJ}$, but rather twenty algebraic equations, fixing a priori the twenty components of $\phi^{IJKL}$ as functions of $B^{IJ}$. The (modified) dynamics is then governed by the whole of \Ref{FE3}. 
As constraints are being removed, one might naively expect additional degrees of freedom, and this is indeed the case. The canonical analysis performed in \cite{Alexandrov} showed the presence of 8 degrees of freedom, with the only condition that the Hessian of $\La(\phi)$ be non-singular. Identifying these degrees of freedom is the goal of the present paper.

Since the original action \Ref{S} is equivalent to general relativity, one might wonder whether the modification is related in any way to the more familiar extension of the Einstein-Hilbert action by adding higher curvature invariants of the metric, with the arbitrary function of $\phi$ playing the role of an arbitrary function of 
$R_{\mu\nu\rho\sigma}$, and with the six extra degrees of freedom related to the ones of these higher derivatives theories \cite{Stelle,Deruelle}.
The answer is negative: as we will show below, the action actually describes a type of \emph{bimetric} gravity, with $\La(\phi)$ related to the interaction potential between the two metrics. The total counting of eight degrees of freedom is consistent with previously known examples of bigravities \cite{Damour}, but the explicit form of the action we obtain is new.
 
These results and the connection with bigravities are rooted in the fact that removing the simplicity constraints, the two metrics naturally present in a non-degenerate $B$ field, see \Ref{ParamGen}, become independent.
Before showing the construction explicitly, let us make a few remarks on this type of modification of general relativity.

\begin{itemize}

\item \emph{BF plus potential} 

As mentioned above, a consequence of the Plebanski constraints in \Ref{S} is to partially fix the large gauge symmetry of BF theory, leaving only diffeomorphisms and gauge transformations. 
The situation is the same in the modified theory: the constraints are replaced by a potential $V(\phi,B)$, which again results in the same partial gauge-fixing -- at least for generic choices of $\La(\phi)$. Hence, each modified theory of gravity corresponds to a specific gauge-fixing of BF theory caused by a potential term, a point discussed in \cite{KrasnovEff}. 
BF theory itself can be recovered for the singular case $\La(\phi) = \d(\phi)$, therefore the class of actions \Ref{Smod1} interpolates between general relativity at constant $\La$, and BF theory at $\La(\phi) = \d(\phi)$.

\item \emph{Not the most general action}

The form \Ref{Smod1} of the action we consider here is motivated by the Hamiltonian analysis performed in \cite{Alexandrov} and the desire to identify the extra degrees of freedom there found. However, this is not the most general action that one can write compatible with the symmetries. In particular, there are two immediate terms that one can add. The first one is the kinetic term $\eps_{IJKL} B^{IJ} \w F^{KL}(\om)$, whose (inverse) coupling constant is usually referred to as the Immirzi parameter. The second is the alternative potential term $\La_2(\phi) \d_{IJKL} B^{IJ} \w B^{KL}$, which is usually not considered in the Plebanski action \Ref{S} because it vanishes on the solutions \Ref{Bsols}.\footnote{However it plays a role if an Immirzi parameter is introduced through the variation of $\phi$, see \cite{PlebImm,Montesinos}.}

\item \emph{Self-dual case}

The additional degrees of freedom are absent when one considers the same modification in the self-dual theory \cite{Krasnov2}. A key difference between the self-dual and the non-chiral actions can be inferred from our earlier decomposition \Ref{S2}. This shows that the non-chiral action is not just the sum of one right- and one left-handed Plebanski actions, due to the mixing terms in $\psi^{ij}$ and $\varphi_0$. The constraints of the non-chiral theory are then not just the sum of 5 and 5 from two separed single-handed actions, but 10 additional ones are present.
Consequently, when we look at the modified theory, more constraints are being lost. This is the origin of the different behaviour under the same type of modification of the two actions, the self-dual and the non-chiral one.
That is, in the non-chiral case it is truly the removal of the constraints ${\bf (1,1)}$ and ${\bf (0,0)}$ which is responsable for the extra degrees of freedom.

This said on the difference, we remark that the two modifications share nonetheless an important characteristic: in both cases, the counting of degrees of freedom is consistent with (the field components parametrized by) the auxiliary fields never becoming dynamical. 

\item \emph{On the quantum theory}

Although this paper deals uniquely with the classical properties of \Ref{Smod1}, let us briefly comment on the quantum theory. First of all, there is no reason a priori to expect an improved UV behaviour -- in fact it may be even worse -- than general relativity. However, 
it was argued in \cite{Krasnov} that the class of theories defined by an arbitrary potential $\La(\phi)$ might be closed under renormalization. Since varying the potential one can interpolate between general relativity and BF theory, 
 one can entertain the rather spectacular possibility that the action has a flow under renormalization that touches only the potential term $\La(\phi)$, evolving it from a constant $\La$ at low energies -- thus describing general relativity with a cosmological constant -- to a conformally invariant (as a matter of fact, even topologically invariant) fixed point $\d(\phi)$ at high energies. Under the additional condition that the flow has a finite dimensional critical surface, this would be a rather intriguing realization of the asymptotic safety scenario. The action \Ref{Smod1} is not in a form easily treatable with conventional renormalization group techniques, but what is at stake seems to us worth the effort of pursuing this direction of studies.

\end{itemize}

\section{From the modified Plebanski action to bi-metric gravity}\label{SecBi}

We are now ready to go back to \Ref{Smod1}, and show how it can be recasted as a bigravity theory.
The structure of the field equations suggests the natural strategy of using \Ref{FE1} and \Ref{FE2} to eliminate respectively $\om$ and $\phi$ (when possible -- see below), and then study the dynamics determined by the equations \Ref{FE3}. In this way, the arbitrary potential $\La(\phi)$ is mapped into an arbitrary potential for $B$, and one is effectively dealing with an action for the 2-form $B$ only.
Finally, using the explicit parametrization \Ref{ParamGen}, the action is recasted into a theory for the two metrics $g_{\mu\nu}$, $\bar g_{\mu\nu}$, and the auxiliary scalars $b^i_a$, $\bar b^i_a$.
The auxiliary fields can in principle be integrated out, and the resulting dynamics described purely in terms of the two metrics, interacting through a given potential. 

When is this construction explicitly possible?
The compatibility condition \Ref{FE1} can be solved uniquely, provided the same non-degeneracy conditions behind \Ref{ParamGen} hold \cite{Deser, Halpern, Bengtsson}. As for \Ref{FE2}, the solution will not be unique in general, leading to additional subsectors. To simplify the analysis, we consider in the rest of this paper the simplest case with a unique solution $\phi(B)$,
\equ\label{La2}
\La(\phi)=\La -  \f{3}{2A} \, {\rm Tr} \, \phi^2. 
\nequ
Here $\La$ is the cosmological constant, $A$ a free parameter with dimensions of a squared mass, 
and ${\rm Tr} \, \phi^2 \equiv \phi_{IJKL} \phi^{IJKL}$.  
We will comment later on more general choices. This quadratic case has a non-singular Hessian, thus the result of \cite{Alexandrov} holds, and we expect eight degrees of freedom.

With the choice \Ref{La2}, the field equations \Ref{FE2} and \Ref{FE3} read
\begin{subequations}\label{newFE}\begin{align}
\label{newFE2}
& B^{IJ}\w B^{KL} = \left(\f1{12} \eps^{IJKL} + \f1A \phi^{IJKL} \right) \, \BsB, \\ \label{newFE3}
& F^{IJ} = \phi^{IJKL} B_{KL} + \left(\f\La6 - \f{1}{4A} {\rm Tr} \, \phi^2\right) \eps^{IJKL} B_{KL}.
\end{align}\end{subequations}
By inspection, it should be clear that solutions of general relativity with vanishing $\phi$ are also solutions of the modified theory, because the modification disappears in that case. Since in general relativity (in Plebanski variables) $\phi$ is on-shell the Weyl tensor (see \cite{Mike, KrasnovPleb, PlebImm}), we conclude that conformally flat spacetimes are still solutions.\footnote{This can change if one includes an Immirzi parameter through the variation of $\phi$ \cite{PlebImm}.} 

To understand the modified theory more in general, we take the strategy outlined above and reformulate \Ref{Smod1} as a bi-gravity action $S(e^I_\mu, \bar e^I_\mu, b^i_a, \bar b^i_a)$ in terms of two metrics and auxiliary scalar fields.
This can be done in three steps: (i) solving the Gauss law to obtain $\om(B)$, (ii) solving \Ref{newFE2} to obtain $\phi(B)$, and (iii) inserting the decomposition \Ref{ParamGen} of the $B$ field. Notice that (i) is relevant to the first (``kinetic'') term of the action only, whereas (ii) to the remaining (``potential'') terms.
We separe our analysis accordingly.

It will be also convenient to introduce an internal metric associated to the triad $b^i_a$, i.e.
\equ\label{defb}
q_{ab} = b^i_a b^j_b \d_{ij},
\nequ 
and similarly $\bar q_{ab}$ for $\bar b^i_a$. These new metrics are also unimodular, and we denote the trace $q \equiv \d^{ab} q_{ab} = m$.
We keep a convention where the $a$ indices are raised and lowered with the identity metric $\d_{ab}$, thus
the inverse of $q_{ab}$ is not $q^{ab}=\d^{ac}\d^{bd}q_{cd}$, but 
\equ
\hat q^{ab} = \hat b^a_i \hat b^b_j \d^{ij} =
\f12 \eps^{acd} \eps^{bef} q_{ce} q_{df}.
\nequ
Notice that $q^{ab} q_{ab} = m^{ij} m_{ij}$.

\subsection{Kinetic term: the effective BF action}

In this section we consider a single-handed SU(2) BF action $\int B^i\w F_i$, and review how its dependence on $B^i$ and $\om^i$ is recasted solely in terms of $b^i_a$ and $e^I_\mu$. 
To do so, one has to solve the compatibility condition ${\rm d}_\om B = 0$ for $\om(B)$. 
This has been known for a generic gauge group for quite some time \cite{Deser, Halpern, Bengtsson}. A particularly useful expression has been recently derived by Freidel \cite{Freidel}, who exploited the decomposition \Ref{BSigma} to give a solution $\om_\eps(b^i_a,\Si^i_\eps(e))$, which carries the familiar spin connection for the tetrad $e_\mu^I$. We report the explicit formula and its derivation in Appendix \ref{AppSF}. The solution is valid for invertible tetrads, and unique.
Restricting to invertible tetrads, one can then integrate out $\om$ in the action using this solution.
The result is the following ``BF effective action'' \cite{Freidel},
\equ\label{SF}
S_{\rm F}(e^I_\mu, q_{ab}) = \f{\eta\eps}4 \int e R_\eps^{ab}(e) (\hat q \d_{ab} - \hat q_{ab}) 
+ \f12 e \n_\eps^\mu q_{ab} C_{\eps \ \mu\nu}^{abcd}(e,q) \n_\eps^\nu q_{cd},
\nequ
where 
\equ\label{Riem}
R^{ab}_\eps(e)=\f12 \Sigma^a_\eps{}_{\mu\nu}(e)\Sigma^b_\eps{}_{\rho\sigma}(e)R_{\mu\nu\rho\sigma}(e)
\nequ 
is the self- or antiself-dual (resp. for $\eps=\pm$) part of the Riemann tensor, $\n_\eps{}_\mu$ is the covariant derivative with respect to the spin connection $\ga_\eps^a{}_b(e)=\eps^a{}_{bc} P^c_\eps{}_{IJ}\om^{IJ}(e)$, and finally
\equ\label{C}
C^{abcd}_{\eps \ \mu\nu}(e,q) \equiv \left(\d^{ad}\d^{bc} - \f12 \d^{ab}\d^{cd} \right) g_{\mu\nu} 
+ \left(\d^{bc} \eps^{ad}{}_g - \hat q^{bc} \eps^{adf} q_{fg} \right) \Sigma_\eps^g{}_{\mu\nu}(e).
\nequ
For completeness and because our notation is slightly different from \cite{Freidel}, we review the derivation of this result in Appendix \ref{AppSF}.
The Freidel BF effective action \Ref{SF} is second order in derivatives, polynomial (of order 5) in $q_{ab}$, and non-polynomial in the tetrad through $R_{ab}(e)$. The dependence of the action on the tetrad is via $\Si(e)$ (or directly via the metric as in $R_{ab}(e)$),
thus it should be truly understood as $S_{\rm F}(\Si^i_\eps(e), q_{ab})$, a functional of 18 independent field components.

Now, notice that setting $q_{ab}=\d_{ab}$, \Ref{SF} reduces to $(\eta\eps/2)e\d_{ab}R^{ab}_\eps(e)=(\eta\eps/2)eR(e)$, namely the Einstein-Hilbert lagrangian. This is to be expected, since $q_{ab}=\d_{ab}$ is precisely the metricity constraint of the self-dual Plebanski theory. 
What is perhaps more surprising about \Ref{SF} is that it shows that BF theory, which is topological, can be formulated in terms of the Riemann tensor \Ref{Riem} of an artificial metric, plus additional terms. The presence of these extra terms restores the shift symmetry $B^i\mapsto B^i + {\rm d}_\om \eta^i$, which is what eliminates any local degrees of freedom present a priori in the metric. Rewriting the action as \Ref{SF} masks the original shift symmetry, but we know that we can first use the SU(2) symmetry to diagonalize the matrix of scalar fields $q_{ab}$, and second fix $q_{ab}=\d_{ab}$ using two components (recall that $q_{ab}$ is unimodular so it has only two independent eigenvalues) of the triple of 1-forms $\eta^i$.
This leaves precisely 10 components in $\eta^i$ that can be used to gauge away the metric.
In other words, every solution of general relativity is also a solution of BF theory in a certain gauge, and any other metric (compatible with the topology of spacetime) is still a solution and can be obtained through a gauge transformation.
We will show this explicitly at the linearized level below.

\subsection{Potential term}
The potential term breaks the shift symmetry reviewed above, thus allowing physical propagation.
With the specific form \Ref{La2}, the potential term in the action has the form
\equ\label{V1}
V(\phi,B) = -\f12 \int \left(\phi_{IJKL} - \f{1}{4A} {\rm Tr}\,\phi^2 \, \eps_{IJKL}\right) B^{IJ} \w B^{KL}+\f13\La \BsB.
\nequ
We can use the field equations \Ref{newFE2} to eliminate $\phi$, and after some trivial algebra,
\equ\label{V2}
V(\phi(B),B) = \f1{24} \int {\BsB}\left[A \left(1- 6 \f{(B^{IJ}\w B^{KL})\,(B_{IJ}\w B_{KL})}{\BsB^2}\right) -4\La\right],
\nequ
with abuse of form notation.
Using the decomposition \Ref{ParamGen} and the metrics $q_{ab}$ and $\bar q_{ab}$ defined as in \Ref{defb}, this gives
\equ\label{Veb}
V(e^I_\mu, \bar e^I_\mu, q_{ab}, \bar q_{ab}) =  \f{A}{12} \int {d^4x}(e q +\bar e\bar q)
\left[1-6\f{e^2 q^{ab}q_{ab} + \bar e^2 \bar q^{ab}\bar q_{ab} + 2 \ell^{ij} \ell_{ij}}{(e q + \bar e\bar q)^2}-\f{4\La}{A}\right],
\nequ
where
\equ\label{ll}
\ell^{ij} \ell_{ij} = \f1{2^6} \eps^{\mu\nu\rho\sigma} \eps^{\alpha\beta\gamma\d} 
q_{ab} \Si^a_{\mu\nu}(e) \Si^b_{\alpha\beta}(e) \bar q_{cd} \bar \Si^c_{\rho\sigma}(\bar e) \bar \Si^d_{\gamma\d}(\bar e).
\nequ

By inspection of the potential, it is clear that \Ref{ll} is the quantity controlling the mixing between the right- and left-handed sectors. We already know that it vanishes whenever $\bar e_\mu^I$ is proportional to $e_\mu^I$. Here we also notice that if the extra scalar fields are fixed to the identity matrix, we have
\eqa\label{ellell}
\ell^{ij} \ell_{ij}(q_{ab}=\bar q_{ab}=\d_{ab}) &=& \f1{16}\eps^{\mu\nu\rho\sigma} \eps^{\alpha\beta\gamma\d}
\left(g_{\mu[\al}g_{\beta]\nu}+\f1{2e}\eps_{\mu\nu\al\beta}\right)
\left(\bar g_{\rho[\ga}\bar g_{\d]\sigma}-\f1{2\bar e}\bar \eps_{\rho\sigma\ga\d}\right)= \nn \\ &=&
\f1{16}(\eps^{\mu\nu\rho\sigma} \eps^{\alpha\beta\gamma\d} 
g_{\mu\alpha} g_{\nu\beta} \bar g_{\rho\gamma} \bar g_{\sigma\d} - 4! e \bar e),
\neqa
where in the first line we used the identity \Ref{ggpluseps} from the Appendix. This will be useful when studying perturbation theory below.

\subsection{Bi-metric gravity}

Let us collect our results. To simplify the notation, we remove the $\scr (\pm)$ subscripts everywhere, with the tacit understanding that all derivatives and curvatures associated to $e_\mu^I$ (resp. $\bar e_\mu^I$) are right-handed (resp. left-handed). 
Adding together the right- and left-handed versions of \Ref{SF} and the potential \Ref{Veb} we obtain 
\eqa\label{Sbimetric}
S(\eta,e^I_\mu, \bar e^I_\mu, q_{ab}, \bar q_{ab}) &=& 
\f14 \int e R^{ab}(e) (\hat q \d_{ab} - \hat q_{ab}) - \f{4}3 q e \La
+ \f12 e \n^\mu q_{ab} C_{\mu\nu}^{abcd} \n^\nu q_{cd} + \nn \\ && 
- \f\eta4 \int \bar e R^{ab}(\bar e) (\hat{\bar q} \d_{ab} - \hat{\bar q}_{ab}) - \f{4}3 \bar q \bar e \La
+ \f12 \bar e \n^\mu \bar q_{ab} C_{\mu\nu}^{abcd} \n^\nu \bar q_{cd} +\nn \\ && 
+\f{A}{12} \int (e q +\bar e\bar q)
\left[1  - 6\f{e^2 q^{ab} q_{ab} + \bar e^2 \bar q^{ab} \bar q_{ab} + 2 \ell^{ij} \ell_{ij}}{(e q + \bar e\bar q)^2}\right].
\neqa
This is our main result.
What we have shown is that 
for invertible tetrads, the modified Plebanski theory is a bi-metric theory of gravity coupled to extra scalar fields $q_{ab}$, $\bar q_{ab}$, and whose action is given by the sum of two Freidel ``effective BF'' actions plus a potential term mixing the right- and left-handed sectors. 
The specific form of the potential comes from the choice \Ref{La2} for $\La(\phi)$. Changing $\La(\phi)$ would affect only the last line of \Ref{Sbimetric}, namely the interaction between the two metrics.\footnote{Notice that both metrics have the same cosmological constant $\La$. Different cosmological constants for the two metrics can be obtained including the second potential term $\La_2(\phi) \d_{IJKL} B^{IJ} \w B^{KL}$ mentioned earlier.}

The presence of the extra scalar fields is the most obvious difference with other bi-metric theories of gravity appeared in the literature. The scalars mediate the interactions between the two metrics, and can not be immediately integrated out since the enter the action non-polynomially. 
At first sight, it looks like the scalar fields $q_{ab}$ and $\bar q_{ab}$ have also acquired a kinetic term, and are thus fully dynamical. However, this Lagrangian is peculiar: its kinetic term is degenerate, as will become apparent below when we study it in perturbation theory. As a consequence, not every field with a kinetic term is actually propagating. Specifically, we will see that the extra scalar fields satisfy algebraic equations, and can therefore be integrated out. In this sense, they are still Lagrange multipliers, as they were in the original BF action.

As already discussed, the physical content of the theory depends on the real parameter $A$: the action has eight propagating degrees of freedom at finite $A$, whereas both limits $A\mapsto \infty$ and $A\mapsto 0$ are singular. In the first case, we see from \Ref{V1} that we are simply removing the modification, thus we go back to the two degrees of freedom of general relativity \cite{Henneaux}. In the second case, we see from \Ref{V2} that it is the whole potential term that vanishes, thus we arrive at BF theory and its zero local degrees of freedom \cite{Blau,Baez}. 
In a sense, the bi-metric theory \Ref{Sbimetric} ``interpolates'' between general relativity and BF theory. 
Since the number of degrees of freedom at finite $A$, eight, matches those of some bigravity theories, it is useful to briefly review them in order to gain some intuition in a simpler case where there are no extra scalars.

Notice also the sign $\eta$, which we recall in the unmodified theory distinguished general relativity ($\eta=-1$) from a topological action ($\eta=1$). Here we see that the previously topological choice $\eta=1$ leads to an action where the two curvature terms have opposite sign. We will focus our analysis on the more stable sector $\eta=-1$.

\subsection{Bigravities}\label{Secbigrav}
Bi-metric theories have appeared long ago in the literature \cite{Isham}, and have been recently looked upon as potential playgrounds for improved quantizations of general relativity and/or describe astrophysical effects without resorting to dark matter/dark energy scenarios \cite{Damour,Blas,Nesti,Hossenfelder,Banados,Milgrom}. The standard action for bigravity has the form
\eqa\label{Sbigravity}
S(e^I_\mu, \bar e^I_\mu) &=& 
\int e [R(e)-2\La]  + \int \bar e [R(\bar e)-2\bar\La]
+ \int (e \bar e)^{1/2} W(\bar g^{\mu\nu} g_{\nu\rho}),
\neqa
where the interaction between the two metrics is parametrized by a scalar potential $W$ of the combination $\bar g^{\mu\nu} g_{\nu\rho}$, the only combination allowed by the common diffeomorphism invariance. The canonical analysis performed by Damour and Kogan \cite{Damour} shows the presence of eight degrees of freedom for a generic $W$. 

The simplest way to identify these degrees of freedom is to take vanishing cosmological constants and to study a perturbative expansion around the same flat backgound for both metric, $g_{\mu\nu}=\d_{\mu\nu}+h_{\mu\nu}$, $\bar g_{\mu\nu}=\d_{\mu\nu}+\bar h_{\mu\nu}$. If we change variables to $h^{\scr{(\pm)}}_{\mu\nu}=(h_{\mu\nu}\pm \bar h_{\mu\nu})\sqrt{2}$, we see that the masslessness of $h^{\scr{(+)}}_{\mu\nu}$ is still protected by the diffeomorphism invariance of the action, whereas the combination $h^{\scr{(-)}}_{\mu\nu}$ is invariant under diffeomorphisms, and can thus acquire a mass term through $W$.
This mass term will in general have the form
\equ\label{MPF}
a h^{\scr{(-)}}_{\mu\nu} h^{\mu\nu}_{\scr{(-)}} + b h_{\scr{(-)}}^2,
\nequ
where the explicit values of the constants $a$ and $b$ depend on the form of $W$. 
It is known since the work of Fierz and Pauli \cite{Fierz} that the choice $a=-b=m^2/4$ is the only one leading to the propagation of five degrees of freedom corresponding to a massive spin 2 particle, whereas any other choice introduces admixtures with an extra scalar mode of negative energy, a ghost signalling the instability of the theory. 

The Fierz-Pauli mass term can be implemented in the bigravity action \Ref{Sbigravity} through an infinite number of potentials $W$, the simplest one being \cite{Isham}
\equ\label{VFP}
W\left(\bar g^{\mu\nu} g_{\nu\rho}\right) = \f{m^2}{4}\left(\f{\bar e}e\right)^{1/2} 
\left(6 \bar g^{\mu\nu} g_{\mu\nu} - (\bar g^{\mu\nu} g_{\mu\nu})^2 + 
\bar g^{\mu\nu} g_{\nu\rho} \bar g^{\rho\sigma} g_{\sigma\mu} -12\right).
\nequ
With this choice, the theory is only propagating a massless and a massive spin-2 particles at the linearized level.
However, it was later shown by Boulware and Deser \cite{Boulware} (see also \cite{Arkani,Creminelli}) that this extra ghost mode is inevitably excited when interactions are included: the canonical analysis of the non-linear completion of the Fierz-Pauli Lagrangian gives six degrees of freedom, the sixth mode being precisely the scalar ghost. The same happens with bigravities: perturbing around the ``doubly flat'' background, the eight degrees of freedom correspond to a massless graviton, a massive spin-2 particle and a scalar ghost. However, it was argued in \cite{Damour} that bigravity theories might be stabilized by the interactions, which would draw the two metrics away from the somewhat pathological doubly flat background, and towards ``bi-cosmological'' configurations. Alternatively, the ghost could also be avoided taking specific potentials \cite{Blas}, or expanding around different backgrounds \cite{Nesti}.

As we will see in the next Section, at least at the linearized level the situation is the same as in bigravity theories. In particular, we can integrate out the auxiliary scalars, and identify the eight degree of freedom with a massless and a massive spin-2 particles, plus a ghost scalar mode.

\section{Perturbative expansion}\label{SecPert}

In this section, we study \Ref{Sbimetric} perturbatively to identify the physical meaning of its eight degrees of freedom.
We take a vanishing cosmological constant, so that the ``doubly flat'' spacetime
\begin{align}\label{flat}
& g_{\mu\nu} = \bar g_{\mu\nu} = \d_{\mu\nu},
& q_{ab} = \bar q_{ab} = \d_{ab},
\end{align}
is an exact solution.
We define the field fluctuations
\begin{align}\label{pert}
& g_{\mu\nu} = \d_{\mu\nu} + h_{\mu\nu}, & \bar g_{\mu\nu} = \d_{\mu\nu} + \bar h_{\mu\nu},
&& q_{ab} = \d_{ab} + \chi_{ab}, && \bar q_{ab} = \d_{ab} +\bar \chi_{ab},
\end{align}
with $\chi_{ab}$ and $\bar \chi_{ab}$ traceless as a consequence of the unimodularity of $q_{ab}$ and $\bar q_{ab}$.

\subsection{Kinetic term}\label{SecExpKin}

To expand the kinetic term, recall that on a flat background the Riemann tensor is given by $R_{\mu\nu\rho\sigma}=\p_\nu\p_{[\rho} h_{\sigma]\mu}-\p_\mu\p_{[\rho} h_{\sigma]\nu} +0(h^2).$
Then using the definition \Ref{Riem}, we find at quadratic order 
\equ\label{Rexp1}
e R_\eps^{ab}(\hat q\d_{ab} - \hat q_{ab}) = 
- \chi_{ab} P^a_\eps{}_{\mu\rho}P^b_\eps{}_{\nu\sigma} \p^\mu \p^\nu h_{\rho\sigma} + 2(eR_\eps)^{(2)},
\nequ
where the projectors $P^a_\eps{}_{\mu\nu}$ are the Plebanski 2-form evaluated on the flat tetrad. Thanks to the expression for the linearized Riemann tensor and the tracelessness of $\chi^{ab}$, the first term equals
\equ
\f12 \chi_{ab} \left(P^a_\eps{}_{\mu\rho}P^b_\eps{}_{\nu\sigma} - \f13 \d^{ab} P^c_\eps{}_{\mu\rho}P^c_\eps{}_{\nu\sigma}\right) R^{\mu\rho\nu\sigma}(h) = \f12 \chi_{ab} C^{ab}(h),
\nequ
where $C^{ab}(h)$ is the (linearized) self-dual part of the Weyl tensor.
The second term in \Ref{Rexp1} is  the linearized Einstein-Hilbert lagrangian
\equ
{\cal L}^{(2)}_{\rm EH}(h_{\mu\nu}) = (eR_\eps)^{(2)} = \f14 \p_\rho h_{\mu\nu} \p^\rho h^{\mu\nu}
-\f12\p_\mu h_{\nu\rho} \p^\nu h^{\mu\rho} + \f12\p_\rho h^{\mu\rho} \p_\mu h - \f14 \p_\mu h \p^\mu h, 
\nequ
independently of $\eps=\pm1$. 
At the same order, we also have
\equ
e D^\mu q_{ab} C^{abcd}_{\eps \ \mu\nu} D^\nu q_{cd} = 
\p^\mu \chi_{ab} \left(\d^{ad}\d^{bc} - \f12 \d^{ab}\d^{cd} \right) \d_{\mu\nu} \p^\nu \chi_{cd} = 
\p_\mu \chi_{ab} \p^\mu \chi^{ab}
\nequ
where in the last equality we used again the tracelessness of $\chi_{ab}$. 

Collecting these results, we get the following linearization of \Ref{SF},
\equ\label{SF3}
S_{\rm F}(h_{\mu\nu},\chi^{ab}) = \f{\eta\eps}2 \int {\cal L}^{(2)}_{\rm EH}(h_{\mu\nu}) 
+\f14 \chi_{ab}C^{ab}(h) - \f14 \chi_{ab} \square \chi^{ab}.
\nequ
As for exact theory \Ref{SF}, the new variables have allowed us to write BF theory as the (linearized) Einstein-Hilbert lagrangian plus extra terms. The latter increase the gauge symmetries of the theory and assure that there are no local degrees of freedom.  In fact, one can check that \Ref{SF3} is invariant under the following transformations,
\begin{align}\label{ShiftLin}
& \d_\xi h_{\mu\nu} = - \xi^{a}{}_{a(\mu\nu)}+\f13 \d_{\mu\nu} \xi, & \d_\xi \chi^{ab} = \xi^{(ab)\mu}{}_{\mu} - \f13 \d^{ab} \xi, 
\end{align}
where
\equ
\xi^{ab}_{\mu\nu} \equiv P^{a}_\eps{}_{\mu\la} \p^\la \eta^{b}_{\nu}, \qquad \xi \equiv \xi^{a}{}_{a}{}^{\mu}{}_{\mu}.
\nequ
This is a linearization of the shift symmetry, and notice that it includes diffeomorphisms, for $\eta^a_\mu = P^a_\eps{}_{\mu\la} \eps^\la$.
Next, the field equations are
\begin{align}\label{FElin}
& E^{\rho\sigma}_{\mu\nu} h_{\rho\sigma} + P^a_\eps{}_{\mu\rho}P^b_\eps{}_{\nu\sigma} \p^\rho \p^\sigma \chi_{ab} = 0,
& \square \chi^{ab} - \f12 C^{ab}(h) = 0,
\end{align}
where $E^{\rho\sigma}_{\mu\nu}$ is the linearized Einstein tensor. Using the SU(2) gauge symmetry to diagonalize $\chi^{ab}$ and then two components of $\eta^i$ to set it to zero, we see that the equations in this gauge imply that both the Ricci and the Weyl tensors vanish, therefore the only solution is the flat metric and there are no local gravitational waves. In particular, the Weyl tensor which is the quantity that in general relativity carries the two physical degrees of freedom, it is also put to zero by a gauge choice.\footnote{In these variables, the absence of local degrees of freedom of 4-dimensional BF theory is reminiscent of what happens to general relativity in the 3-dimensional case, which in fact is directly equivalent to BF theory in the sector of invertible triads.}

\subsection{Modified self-dual theory}

Before continuing with the linearization of our action \Ref{Sbimetric}, let us indulge a moment longer on just the self-dual BF action linearized above.
Following \cite{Freidel}, we introduce the traceless and transverse tensor
\equ\label{chimunu}
\chi_{\mu\nu} =  P^a_\eps{}_{\mu\rho}P^b_\eps{}_{\nu\sigma} \f{\p^\rho \p^\sigma}{\square} \chi_{ab},
\nequ
which is invariant under diffeomorphisms. This allows us to rewrite the action 
\Ref{SF3} in the compact form
\equ\label{SF2}
S_{\rm F}(h_{\mu\nu},\chi^{ab}) = \f{\eta\eps}2 \int {\cal L}^{(2)}_{\rm EH}(h_{\mu\nu}) 
-\f12\chi_{\mu\nu}\square h^{\mu\nu} - \f14 \chi_{\mu\nu} \square \chi^{\mu\nu} \equiv 
\f{\eta\eps}2 \int {\cal L}^{(2)}_{\rm EH}\Big(h_{\mu\nu}+\chi_{\mu\nu}\Big).
\nequ

Consider now the addition of a mass term $m^2 \chi^{ab} \chi_{ab} \equiv m^2 \chi^{\mu\nu} \chi_{\mu\nu}$ for the scalars. This is precisely the type of effect that one has in the modified self-dual theory (cf. the potential term of \Ref{Sbimetric} for $\bar e^I_\mu = \bar q^{ab} = 0$). This mass term breaks the shift symmetry \Ref{ShiftLin}. In this case one can introduce a shifted field $H_{\mu\nu} \equiv h_{\mu\nu} + \chi_{\mu\nu}$ and make the non-singular change of variables $(h_{\mu\nu},\chi^{ab}) \mapsto (H_{\mu\nu},\chi_{\mu\nu})$, in terms of which the action reads
\equ\label{SF4}
S_{\rm F}(H_{\mu\nu},\chi_{\mu\nu}) = \f{\eta\eps}2 \int {\cal L}^{(2)}_{\rm EH}(H_{\mu\nu}) + m^2 \chi^{\mu\nu} \chi_{\mu\nu}.
\nequ
Now the auxiliary scalars satisfy the algebraic field equations $\chi^{\mu\nu}=0$, and can be thus trivially integrated out. We then see that the shifted field $H_{\mu\nu}$ propagates the two degrees of freedom of a massless spin-2 particle, precisely as an ordinary graviton $h_{\mu\nu}$. This is the mechanism that makes the modified self-dual Plebanski theory only propagate two degrees of freedom. See \cite{Freidel, KrasnovEff} for a discussion of this.
In other words, the action  \Ref{SF4} has a degenerate kinetic term, and some of the fields only satisfy algebraic equations. This result can be interpreted in the light of the fact that prior to adding the potential term, the $B$ field in the BF action is just a Lagrange multiplier. What \Ref{SF4} shows is that even when the potential term is present, some components of the $B$ field still behave like Lagrange multipliers and satisfy non-dynamical algebraic equations.

\subsection{Potential term}

Going back to the full non-chiral theory, let us rewrite the potential term in the convenient form
\equ\label{Vpert}
V = \f{A}{2} \int (e q +\bar e\bar q)^{-1}
\left[\f16(e q + \bar e\bar q)^2  - e^2 q^{ab} q_{ab} - \bar e^2 \bar q^{ab} \bar q_{ab} - 2 \ell^{ij} \ell_{ij} \right].
\nequ
This vanishes on the bi-flat solution \Ref{flat}. To expand it at second order, we use the familiar formula 
\equ
e = 1+\f12 h +\f14 (\f12h^2-h^{\mu\nu}h_{\mu\nu})
\nequ
for the determinant, and compute
\begin{align}
&\f16(eq + \bar e \bar q)^2 = 6 \left( 1+\f12(h+\bar h)+
\f1{16}\left(3 h^2 +3 \bar h^2 +2h \bar h\right)-\f14 h^{\mu\nu}h_{\mu\nu} - \f14\bar h^{\mu\nu}\bar h_{\mu\nu}\right),
\nn\\ \nn
&e^2q^{ab}q_{ab} = 3\left(1+ h + \f12( h^2 -h^{\mu\nu}h_{\mu\nu})\right) + \chi^{ab} \chi_{ab}.
\end{align}
Using these expressions, we have
\equ\label{int1}
\f16(e q + \bar e\bar q)^2  - e^2 q^{ab} q_{ab} - \bar e^2 \bar q^{ab} \bar q_{ab} \simeq
-\f38 (h-\bar h)^2 - \chi^{ab} \chi_{ab} - \bar\chi^{ab} \bar\chi_{ab}.
\nequ

It remains to evaluate the term $-2\ell^{ij}\ell_{ij}$. Its zeroth and first orders vanish due to the orthogonality of the Plebanski 2-forms $\Si$ and $\bar \Si$ when evaluated on the same metric, and the only non-vanishing second order contribution comes from its value \Ref{ellell} at $q_{ab}=\bar q_{ab}=\d_{ab}$, which gives
\equ\label{int2}
-2\ell^{ij}\ell_{ij} \simeq -\f18(\eps^{\mu\nu\rho\sigma} \eps^{\alpha\beta\gamma\d} 
g_{\mu\alpha} g_{\nu\beta} \bar g_{\rho\gamma} \bar g_{\sigma\d} - 4! e \bar e) \simeq 
-\f12( h_{\mu\nu}- \bar h_{\mu\nu})^2 +\f18 (h-\bar h)^2.
\nequ
Since the numerator of \Ref{Vpert} has no zeroth nor first order terms, the denominator only contributes $(e q +\bar e\bar q)^{-1}=1/6$. The expansion thus yields
\equ\label{Vexp3}
V^{(2)} = -\f{A}{2} \int \f14 (h-\bar h)^2 +\f12( h_{\mu\nu}- \bar h_{\mu\nu})^2 + \chi^{ab} \chi_{ab} + \bar\chi^{ab} \bar\chi_{ab}.
\nequ

We see that the potential is giving a mass to the auxiliary scalar fields and, as expected from the previous discussion, also to the diffeomorphic-invariant combination of gravitons.

\subsection{Mass eigenstates}

Putting the expansions \Ref{SF2} and \Ref{Vexp3} together and introducing the projections \Ref{chimunu} as before, we find
\eqa
S(h_{\mu\nu}, \bar h_{\mu\nu}, \chi^{ab}, \bar \chi^{ab}) &=& 
\f{1}2 \int {\cal L}^{(2)}_{\rm EH}\Big(h_{\mu\nu} + \chi_{\mu\nu} \Big) -\eta
{\cal L}^{(2)}_{\rm EH}\Big( \bar h_{\mu\nu} + \bar \chi_{\mu\nu} \Big) 
\nn \\ && -
\f{A}{2} \int  \f14 (h-\bar h)^2 +\f12( h_{\mu\nu}- \bar h_{\mu\nu})^2 + \chi^{\mu\nu} \chi_{\mu\nu} + \bar\chi^{\mu\nu} \bar\chi_{\mu\nu}.
\neqa
From now on we restrict attention to the sector $\eta=-1$, which we recall corresponded in the unmodified action to general relativity . 
The mass term for the gravitons can be diagonalized introducing the linear combinations 
$$
h^{\scr{(\pm)}}_{\mu\nu} = \f1{\sqrt{2}}(h_{\mu\nu}\pm \bar h_{\mu\nu}), \qquad 
\chi^{\scr{(\pm)}}_{\mu\nu} = \f1{\sqrt{2}}(\chi_{\mu\nu}\pm \bar \chi_{\mu\nu}).
$$
Since this transformation preserves the kinetic term, we get
\eqa
S(h^{\scr{(+)}}_{\mu\nu}, h^{\scr{(-)}}_{\mu\nu}, \chi^{\scr{(+)}}_{\mu\nu}, \chi^{\scr{(-)}}_{\mu\nu}) &=& 
\f{1}2 \int {\cal L}^{(2)}_{\rm EH}\Big(h^{\scr{(+)}}_{\mu\nu} + \chi^{\scr{(+)}}_{\mu\nu} \Big) +
{\cal L}^{(2)}_{\rm EH}\Big(h^{\scr{(-)}}_{\mu\nu}+\chi^{\scr{(-)}}_{\mu\nu} \Big) 
\nn \\ && -
\f{A}{2} \int  \f12 h_{\scr{(-)}}^2 + h_{\scr{(-)}}^{\mu\nu}h^{\scr{(-)}}_{\mu\nu} + \chi_{\scr{(+)}}^{\mu\nu} \chi^{\scr{(+)}}_{\mu\nu} + \chi_{\scr{(-)}}^{\mu\nu} \chi^{\scr{(-)}}_{\mu\nu}.
\neqa

As we did above in the self-dual sector, we can introduce the shifted fields $H^{\scr{(\pm)}}_{\mu\nu} \equiv h^{\scr{(\pm)}}_{\mu\nu} + \chi^{\scr{(\pm)}}_{\mu\nu}$ and change variables, obtaining
\eqa
S(H^{\scr{(+)}}_{\mu\nu}, H^{\scr{(-)}}_{\mu\nu}, \chi^{\scr{(+)}}_{\mu\nu}, \chi^{\scr{(-)}}_{\mu\nu}) &=& \f{1}2 \int {\cal L}^{(2)}_{\rm EH}(H^{\scr{(+)}}_{\mu\nu}) + {\cal L}^{(2)}_{\rm EH}(H^{\scr{(-)}}_{\mu\nu}) 
+ \\ \nn && -
\f{A}2 \int \f12 H_{\scr{(-)}}^2 + H_{\scr{(-)}}^{\mu\nu} H^{\scr{(-)}}_{\mu\nu} 
-2 H_{\scr{(-)}}^{\mu\nu}\chi^{\scr{(-)}}_{\mu\nu} + \chi_{\scr{(+)}}^{\mu\nu} \chi^{\scr{(+)}}_{\mu\nu} 
+2 \chi_{\scr{(-)}}^{\mu\nu} \chi^{\scr{(-)}}_{\mu\nu}.
\neqa
This is a non local field redefinition, but it has the advantage that now the auxiliary fields satisfy the algebraic equations $\chi_{\scr{(+)}}^{\mu\nu}=0$ and $\chi_{\scr{(-)}}^{\mu\nu} = H_{\scr{(-)}}^{\mu\nu} / 2$. We can thus trivially integrate them out and describe this quadratic lagrangian purely in terms of the two shifted gravitons $H^{\scr{(\pm)}}_{\mu\nu}$.
 Specifically, $H^{\scr{(+)}}_{\mu\nu}$ still propagates only two degrees of freedom as it was the case for the modified self-dual theory (see Section \ref{SecExpKin}), whereas $H^{\scr{(-)}}_{\mu\nu}$ acquires a mass term: its field equation is 
\equ\label{massiveFE}
E^{\rho\sigma}_{\mu\nu}H^{\scr{(-)}}_{\rho\sigma} -{A} \left( H^{\scr{(-)}}_{\mu\nu} + H^{\scr{(-)}} \d_{\mu\nu}\right). 
\nequ
Notice that the mass term is not of the Fierz-Pauli type, thus these are field equations propagating both a massive spin-2 particle and a massive scalar.

\subsection{Lorentzian signature and reality conditions}\label{SecLor}

In order to properly talk about degrees of freedom, we need to move to the physical Lorentzian signature. The formulas can be easily adapted (the main differences stemming from the $i$ factors now present in the right- and left-handed projectors), but there is a caveat. The isomorphism of $\so{3,1}$ into right- and left-handed sectors, which is the core of our construction, requires the complexification of the algebra. Hence, the key field decomposition \Ref{ParamGen} now requires $B^{IJ}$ to be complex a priori, and suitable reality conditions have to be included in order to define the physical sector of the theory.
The difficulty here lies in the fact that different reality conditions can be envisaged. In particular, it risks to be premature to discuss reality conditions before matter coupling, a crucial aspects which is still to be fully addressed in this modified non-chiral theory.\footnote{For instance, a natural way to include matter in the theory is through grand unification, which was the original reason for considering this modification of the non-chiral Plebanski action \cite{Lee,noi1}. However, precisely this idea shows an example of how the reality conditions depend on what one is trying to do. For instance, the same modified non-chiral action considered here could be taken as a toy gravity-gauge unification, a special case of the more general setting investigated in \cite{Torres}. To do that, one takes the right-handed sector alone to describe gravity, and expands instead the left-handed sector around a degenerate background, thus keeping the connection as the fundamental variable. In this case, one would need different reality conditions to ensure the physical interpretation of such attempt to gravity-gauge unification, naturally associated to the new fundamental variables. We thank K. Krasnov for pointing this out to us.}
At the present stage, the simplest thing one can do is to define the Lorentzian theory with all fields complex, and impose the same reality conditions used in the self-dual theory \cite{Capo2,KrasnovEff}, separately on both right- and left-handed sectors. This guarantees that $g_{\mu\nu}$ and $\bar g_{\mu\nu}$ are real Lorentzian metrics. If we do so, the above analysis can be straighforwardly adapted, and we conclude that the modified Plebanski theory with action \Ref{Sbimetric} is propagating a massless and a massive spin-2 particles, plus a scalar mode.

We need at this point to discuss the (in)stability of the perturbative expansion around the bi-flat solution. The massive spin-2 particle can be made stable taking the appropriate sign of $A$.\footnote{Qualitatively different behaviours in the branches $A>0$ and $A<0$ are to be expected, and can be found also in the modified self-dual theory, see for instance \cite{Ishibashi}.}
On the other hand, it is well known from the study of massive gravity that given the structure of \Ref{massiveFE}, there is no choice of $A$ that would make the scalar mode stable.
This means that the chosen potential \Ref{La2} gives a bigravity theory which is perturbatively unstable around the doubly flat solution. As mentioned above in Section \ref{Secbigrav}, this situation is rather generic in bigravity theories.
On the other hand, the stability around a different background (cf. \cite{Damour, Nesti, Banados}), or at the non-perturbative level \cite{Arkani,Creminelli}, or with a different potential $\La(\phi)$ (cf. \cite{Blas}), remains an open issue to investigate. 
In the latter perspective, it was already suggested in \cite{Alexandrov} that one might look for different profiles of $\La(\phi)$ with the goal of reducing the extra degrees of freedom. Let us add here that this idea can be further generalized, since as we remarked above in Section \ref{SecMod}, the action we have been considering is not the most generic one that can be written down.

Finally, it would also be interesting to investigate different reality conditions, as they might affect our conclusions. In particular, notice that having relaxed the simplicity constraints, the reality conditions considered above do not imply that $B^{IJ}$ is real, as it was the case with the standard solution \Ref{Bsols}. Thus although we have identified eight degrees of freedom, strictly speaking these are not the ones discussed for Lorentzian signature in \cite{Alexandrov}, since there both signatures are defined using real fields. To stick with \cite{Alexandrov} in the Lorentzian case, we need different reality conditions, namely imposing $B^{IJ}$ to be real. This immediately implies that $B^{i}_{\scr -} \equiv (B^{i}_{\scr +}){}^*$, where ${}^*$ stands for complex conjugate. An implementation is to take $\bar b^i_a = (b^i_a)^*$ and $\bar g_{\mu\nu} =(g_{\mu\nu})^*$, with the independent fields $(b^i_a, g_{\mu\nu})$ complex. Therefore, we still have a bimetric theory, although in a different flavour, where the two metrics are the real and imaginary parts of the single complex metric emerging after imposing the reality conditions. This alternative construction, which is the one truly corresponding to \cite{Alexandrov} for Lorentzian signature, should also be explored. However, as said above, we feel that a discussion of the reality conditions should parallel the one of matter coupling, thus we postpone both to further studies.

\section{Conclusions}\label{SecConcl}

The modified Plebanski actions, introduced by Krasnov and Smolin, and related to earlier work by Capovilla, Bengtsson and Peldan, are proving to be an interesting arena to deepen our understanding of this formalism for gravity, in which the fundamental field is a 2-form $B$, and the metric only a derived quantity. In our view, an important lesson is the ``watering down'' of the role of the constraints present in the original actions. First of all, the role of the constraints is not to introduce a metric, since a metric is already present, but rather to single it out among the various components of the $B$ field. This was already known, and our review at the beginning of the paper was meant at stressing this aspect. But what is more important, the study of the modified actions shows that the true mechanism at play is a symmetry breaking, as discussed in \cite{KrasnovEff}: one starts with the topological BF action, and adds a term that breaks the initial shift symmetry down to diffeomorphisms only.  When this happens, some of the components of $B$ remain Lagrange multipliers, whereas others become dynamical. The dynamical ones can be encoded in one metric (or two, depending on the gauge group). 
This is the key mechanism, and the symmetry breaking term does not need to be a constraint for $B$, it can be an arbitrary potential term.
 Finally, the specific type of symmetry breaking term determines the dynamics followed by the metric. It can be made arbitrarily close to GR, the exact correspondence being obtained for the singular case in which the potential becomes a constraint.

An interesting aspect of the modification is that it differentiates \emph{qualitatively} the self-dual and non-chiral Plebanski actions, which are equivalent (and equivalent in the non-degenerate sector to general relativity) in the original formulation. In particular, the non-chiral action acquires six additional propagating degrees of freedom \cite{Alexandrov}. In this paper, we studied the modified non-chiral theory with the aim of understanding the origin of this different behaviour, and identifying the extra degrees of freedom. To that end, we focused on the simplest form of the potential. Our first result was to show that the modified action can be recasted in a bi-metric theory of gravity plus auxiliary scalar fields. Our second result was to perform a perturbative expansion around  
the bi-flat background, which is still an exact solution of the modified theory, and identify the eight degrees of freedom in terms of a massless and a massive spin-2 particles, plus a scalar mode. 

Our construction highlights the origin of the different behaviour of the two actions.
In fact, the key to our result is the fact that the non-chiral Plebanski action is naturally a theory of two metrics. It is only the presence of the constraints that imposes these two metrics to be proportional to each other, thus reducing the theory to a single propagating graviton. Once the constraints are traded for a potential, both metrics can independently propagate. Then the fact that two propagating metrics give rise to a massive graviton plus a scalar mode is simply a consequence of the fact that there is a single invariance under diffeomorphism at play \cite{Damour}. 

The main difference with usual bi-gravity theories is the presence of the extra scalar fields.
A subtle mechanism is at play in keeping these non-metric components non-dynamical: after a certain field redefinition, they satisfy algebraic equations and can thus be integrated out. In a sense, they are still Lagrange multipliers as in the original BF action. The procedure is subtle due to the tensorial structure of the fields, and can only be performed order by order in perturbation theory. It is not clear to us if and how the relevant field redefinition can be performed at the level of the full non-perturbative action.

Our results also show that the perturbative expansion around the doubly flat spacetime is unstable for any value of the free parameter in the modified theory. This is the same instability found in non-linear massive gravity and in bigravity theories with eight degrees of freedom. The stability around different backgrounds, or of actions with a different potential, remains an open issue to investigate.

Finally, our results depend on a specific choice of reality conditions, which are necessary to deal with the Lorentzian signature in our construction. Further investigations of the reality conditions, together with an analysis of matter coupling, are in our view the most pressing open issues of this modified gravity theory, which we hope to come back to in future studies.

Concluding, bi-metric theories of gravity are an interesting playground for alternative explainations of the present astrophysical/cosmological puzzles. We hope to have shown with this paper that the Plebanski formalism for general relativity, based on the use of a 2-form as the fundamental field, as opposed to the metric, provides one such playground in a rather natural way.

\section*{Acknowledgements}
The author is indebted to Lee Smolin, with whom this project was started. Useful discussions with D. Beke, C. Skordis, L. Freidel, K. Krasnov, D. Benedetti and F. Nesti are gratefully acknowledged.

\appendix

\section{Algebraic notation and conventions}\label{AppN}

In this Appendix we collect our conventions, and some useful formulas. The isomorphism $\mathfrak{so(4)}\cong \mathfrak{su(2)}\oplus \mathfrak{su(2)}$ is realized by the right- and left-handed projectors
\equ
P_{\eps}^{IJ}{}_{KL} = \f\eps2\left(\d^{IJ}_{KL} + \f\eps2 \eps^{IJ}_{KL}\right),
\nequ
with $\eps=\pm$. 
It is convenient to use indices $i=1,2,3$ for the two $\mathfrak{su(2)}$ algebras. This can be done defining new tensors
\equ
P_{\eps}^{IJ}{}_i = 2 P_{\eps}^{IJ}{}_{0i} = \eps\d^{IJ}_{0i} + \f12 \eps^{IJ}_{0i},
\nequ
for $\eps=\pm$, normalized so to have 
\equ
\d^{ij} P_{ \eps}^{IJ}{}_i P_{ \eps}^{KL}{}_j = \eps P_{ \eps}^{IJKL},
\qquad \d_{IJKL} P_{ \eps}^{IJ}{}_i P_{ \eps}^{KL}{}_j = \d_{ij},
\qquad \f12 \eps_{IJKL} P_{ \eps}^{IJ}{}_i P_{ \eps}^{KL}{}_j = \pm\d_{ij}.
\nequ

We take $\eps^{\mu\nu\rho\sigma}$ to be the completely antysimmetric tensor density, with $\eps^{0123}=1$. We define 
$$
\eps_{\mu\nu\rho\sigma} = g_{\mu\al} g_{\nu\beta} g_{\rho\gamma} g_{\sigma\d} \eps^{\al\be\gamma\d},
\qquad \eps^{\mu\nu\rho\sigma} \eps_{\mu\nu\rho\sigma} = 4! g.$$

The Plebanski 2-form coincides with \Ref{Pi} projected along the tetrad,
\equ
\Si^i_\eps{}_{\mu\nu}(e) \equiv 2 \eps e_{[\mu}^0 e_{\nu]}^i + e^i{}_{jk} e^j_\mu e^k_\nu \equiv 2 P^i_\eps{}_{\mu\nu}.
\nequ
The right-handed (resp. left-handed) Plebanski form is also self-dual (resp. antiself-dual) with respect to its tetrad, i.e.
\equ
P^{\mu\nu}_\eps{}_{\rho\sigma} \Si^i_\eps{}_{\mu\nu}(e) = \Si^i_\eps{}_{\rho\sigma}(e), \qquad 
P^{\mu\nu}_\eps{}_{\rho\sigma} \equiv \f\eps2\left(\d^{\mu\nu}_{\rho\sigma}+\f\eps{2e}\eps^{\mu\nu}{}_{\rho\sigma}\right).
\nequ
Furthermore, we have
\begin{align}
&\Sigma^i_\eps(e) \w \Sigma_\eps^j(e) = \eps \, 2 e \d^{ij} \, d^4x, \\ & \label{ggpluseps}
\f12 \d_{ij} \Si^i_\eps{}_{\mu\nu} \Si^j_\eps{}_{\rho\sigma} = 
g_{\mu[\rho} g_{\sigma]\nu}+\f\eps{2e}\eps_{\mu\nu\rho\sigma},
\\ & 
\label{double}
\Si^i_\eps{}_{\mu\nu} \Si^j_{\eps}{}_{\rho\sigma} g^{\nu\sigma} = \d^{ij} g_{\mu\rho} + \eps^{ij}{}_l \Si^l_\eps{}_{\mu\rho},
\\ & 
\Si^i_\eps{}_{\mu\nu} \Si^j_\eps{}^{\mu\rho} \Si^k_\eps{}_{\rho\sigma} =
\d^{ij} \Si^k_\eps{}_{\nu\sigma} - \d^{ik} \Si^j_\eps{}_{\nu\sigma} + \d^{jk} \Si^i_\eps{}_{\nu\sigma}
-\eps^{ijk} g_{\nu\sigma},
\label{triple}
\end{align}
which can all be easily checked by direct computation.

\section{Review of the Freidel effective action}\label{AppSF}
In this Appendix we review Freidel's construction of the $\su{2}$ BF effective action \cite{Freidel}. We do so to make the paper self-contained, but also to fix the notation (slightly different from \cite{Freidel}) and numerical factors.
The construction is somewhat lenghty, and it is convenient to split it in two steps. In the first, we solve the Gauss law $\rd_\om B^i=0$, and express $\om$ in terms of the parametrization 
\equ\label{Bapp}
B^i_\eps = \eta b^i_a \Si^a_\eps(e). 
\nequ
In the second step, we insert the result in 
\equ\label{Sapp}
S(B^i, \om^i{}_j) = \int B^i\w F_i(\om)
\nequ 
and evaluate the effective action $S(\eta,b,\Si_\eps)$.

\subsection{Solving the Gauss law}

We now come to the Gauss law
\equ\label{Glaw}
\rd_\om B^i=\rd B^i + \om^i{}_j \w B^j = 0.
\nequ 
Solutions $\om=\om(B)$ have been known for a long time \cite{Deser,Halpern,Bengtsson}.
The novelty introduced by Freidel is to solve it in terms of the parametrization \Ref{Bapp}. To do so, we insert \Ref{Bapp} into \Ref{Glaw}. As the $\eta$ is irrelevant, we drop it in this subsection.
We have
\eqa
0 &=& \rd(b^i_a \Si^a_\eps) + b^j_a \om^i{}_j \w \Si^a_\eps =
b^i_a \hat b^a_j \rd b^j_c \w \Si^c_\eps +b^i_a {\rm d}\Si^a_\eps + b^i_a b^j_c \hat b^a_k \om^k{}_j \w \Si^c_\eps = \nn\\ 
&=& b^i_a \left[\rd\Si^a_\eps + 
\left( \hat b^a_j \rd b^j_c +\hat b^a_j\om^j{}_k b^k_c \right)\w\Si^c_\eps \right]. \nn
\neqa
The quantity in the round bracket defines a new connection, which we denote $A^a{}_b$: 
\equ\label{defA}
A^a{}_b \equiv \hat b^a_j \rd b^j_b +\hat b^a_j\om^j{}_k b^k_b = \hat b^a_j \rd_\om b^j_b.
\nequ
Notice that this is the connection compatible with the metric $q_{ab}$, since
\eqa
0 &=& \rd_\om \d_{ij} = \rd_\om(\hat b^a_i \hat b^b_j q_{ab} ) = 
q_{ab} \hat b^b_j \rd_\om \hat b^a_i + q_{ab} \hat b^a_i \rd_\om \hat b^b_j
\hat b^a_i \hat b^b_j \rd q_{ab} = \nn\\ &=&
\hat b^a_i \hat b^b_j \left[\rd q_{ab} - A^c{}_a q_{cb}- A^c{}_b q_{ac} \right] =   \hat b^a_i \hat b^b_j \rd_A q_{ab}.\nn
\neqa
Therefore, solving \Ref{Glaw} amounts to solving 
\equ\label{Glaw1}
\rd_A\Si^a_\eps(e)=0, \qquad \rd_A q_{ab}=0.
\nequ

At this point we assume that the tetrad is invertible, and introduce the right- and left-handed spin connections 
\equ\label{gadef}
\ga^a_\eps{}_b(e) = \eps^a{}_{bc} P_\eps^c{}_{IJ} e^I_\nu \nabla_\mu e^{\nu J}.
\nequ
A direct computation shows that $\rd_{\ga_\eps} \Si^a_\eps$ vanishes. 
To avoid clogging the notation, we remove from now on the subscript $_\eps$ from all the connections: whether it is a right- or left-handed connection should always be clear from the context.
If we write
$A^a{}_b \equiv \ga^a{}_b(e) + \rho^a{}_b$ in terms of the spin connection and an unknown one-form $\rho^a{}_b$, \Ref{Glaw1} read
\equ\label{Glaw2}
\rho^a{}_b\w \Si^b_\eps = 0, \qquad \rd_\ga q_{ab} = 2 \rho^c{}_{(a} q_{b)c}.
\nequ
To solve these equations, we define $\tl \rho_{ab} = q_{ac} \rho^c{}_b = \tl\rho_{(ab)} + \tl \rho_{[ab]}$, 
$\tl \rho_{[ab]}\equiv \eps_{abc} \tl \rho^c$. From the second equation in \Ref{Glaw2} we immediately read off $\tl\rho_{(ab)}=\rd_\ga q_{ab}/2$. The first equation then gives
\equ\nn
\left(\n_\nu q_{cb} + 2\eps_{cbd} \tl\rho^d_\nu \right) \Si^b_\eps{}^{\mu\nu}=0,
\nequ
where $\n_\mu$ indicates the covariant derivative with respect to $\ga^a{}_b(e)$, and we used $\eps^{\mu\nu\rho\sigma}\Si^i_\eps{}_{\rho\sigma}=2e \eps \Si^i_\eps{}^{\mu\nu}$. 
To make $\tl \rho^a$ explicit, we contract the above equaion with $\Si^c_\eps{}_{\mu\rho} \Si^f_\eps{}^{\rho\sigma}$ and use \Ref{triple}. After some algebra, one gets
\equ\label{rhotl}
\tl\rho^a_\sigma = \f12\left[\d^{a(b}\Si^{c)}_\eps{}_{\sigma\la}-\f12\d^{bc}\Si^{a}_\eps{}_{\sigma\la} \right]\n^\la q_{bc}.
\nequ

Hence,
\equ\label{rhosol}
\rho^a{}_b = \hat q^{ac} \tl \rho{}_{cb}, \qquad
\tl\rho_{cb} = \f12 \rd_{\ga} q_{cb} + \eps_{cbd}\tl \rho^d
\nequ 
with $\tl\rho^d$ given by \Ref{rhotl}, and
\equ\label{obs}
\om^i{}_j(b^i_a,\Si^i_\eps)=b^i_a \rd_A \hat b^a_j = b^i_a \rd_\ga \hat b^a_j + 
b^i_a \hat b^b_j \hat q^{ac} \tl\rho_{cb}.
\nequ
We can now use this expression to evaluate the $\su{2}$ curvature $F^i$, appearing in the action \Ref{Sapp}. Recall that this is related to the connection by $F^i(\om)\equiv \f12\eps^i_{jk}(\rd \om^{jk} + \om^j{}_l\w \om^{lk})$; using the expression \Ref{obs}, it is straighforward to see that
\begin{align}\label{Feff}
& F^i(\om(b,\Si_\eps))=\f12 \eps^i{}_{jk}b^j_a\hat b^{ck} F^a{}_b(A_\eps), \\ &
F^a{}_b(A) = F^a{}_b(\ga)+\rd_{\ga} \rho^a{}_b + \rho^a{}_c\w \rho^c{}_b.\label{Feff2}
\end{align}

\subsection{Evaluating the effective action}
Inserting \Ref{Bapp} and \Ref{Feff} in \Ref{Sapp} we get
\equ
S(\eta, b^i_a, \Si^a_\eps) = \f\eta2 \int \eps_{ijk}b^i_a b^j_b\hat b^{ck} \Si^a_\eps \w F^b{}_c(A_\eps)
=\f\eta2 \int \eps_{abd} \hat q^{cd} \Si^a_\eps \w F^b{}_c(A_\eps).
\nequ
This action has three contributions given by the three terms of $F(A)$ in \Ref{Feff2}. Let us start with the first one,
\equ
\eps_{abd} \hat q^{cd} \Si^a_\eps \w F^b{}_c(\ga_\eps)= 
\eps_{abd} \eps^b{}_{ce} \hat q^{cd} \Si^a_\eps \w F^e(\ga_\eps)= (\hat q \d_{ae}-\hat q_{ae}) \Si^a_\eps \w F^e(\ga_\eps)
\nequ
where we used $F^{ab}=\eps^{ab}{}_c F^c$. Thanks to \Ref{gadef} and Cartan's second structure equation \Ref{C2}, we have
\equ
F^a_{\mu\nu}(\ga(e))=P^a_{\eps}{}_{IJ} F^{IJ}_{\mu\nu}(\om(e)) = \f12 \Si^a_\eps{}_{\rho\sigma}(e) R^{\rho\sigma}{}_{\mu\nu}(e).
\nequ
From this it follows that 
\equ\label{uno}
\eps \f{e}2 R^{ae}_\eps \equiv \Si^a_\eps \w F^e(\ga_\eps) = \f18 \eps^{\mu\nu\rho\sigma} \Si^a_\eps{}_{\mu\nu} \Si^e_\eps{}_{\la\tau} R^{\la\tau}{}_{\rho\sigma} = \eps \f{e}4 \Si^a_\eps{}_{\mu\nu} \Si^e_\eps{}_{\rho\sigma} R^{\mu\nu\rho\sigma},
\nequ
which gives us the first term of \Ref{SF} in the main text.
The quantity $R^{ab}_\eps$ here defined is the self-dual part of the Riemann tensor. In particular, using \Ref{ggpluseps} we have 
\equ\label{contraction}
\d_{ae} \Si^a_\eps \w F^e(\ga_\eps) = \f\eps2 e R(e)
\nequ
where $R(e)$ is the Ricci scalar.

The second term is
\equ\label{second}
\eps_{abd} \hat q^{cd} \Si^a_\eps \w {\rd}_{\ga} \rho^b{}_c= 
\eps_{abd} \Si^a_\eps \w \rho^b{}_c \w {\rd}_{\ga} \hat q^{cd}= 
- \eps_{abd} \Si^a_\eps \w \rho^b{}_c \w (\rho^{c}{}_e \hat q^{de}+\rho^{d}{}_e \hat q^{ce}), 
\nequ
where in the first step we integrated by parts, and in the second we used the compatibility condition $\rd_A q_{ab}=0$.

When we add the third term from \Ref{Feff2} we get exactly the first term of the square bracket in \Ref{second} above, but with opposite sign. Therefore these two contributions cancel, and we are left with 
\equ\label{SFApp2}
S(\eta, b^i_a, \Si^a_\eps) = \f\eta4 \int \eps e R^{ae}_\eps (\hat q \d_{ae}-\hat q_{ae}) - 
2\eps_{abd} \hat q^{ce} \Si^a_\eps \w \rho^b{}_c \w \rho^{d}{}_e.
\nequ
In the remaining of this section we evaluate the second term in \Ref{SF2} using the explicit solution \Ref{rhosol}.
We have
\eqa\label{pippo}
&& -\f\eta2\eps_{abd} \hat q^{ce} \Si^a_\eps \w \rho^b{}_c \w \rho^{d}{}_e =
-\f\eta2\eps_{abd} \hat q^{ce} \hat q^{bf} \hat q^{dg} \Si^a_\eps \w \tl \rho_{fc} \w \tl \rho_{ge} =
\nn \\ && \qquad = -\f\eta8 \eps^{hfg}q_{ah}\hat q^{ce}
\Si^a_\eps \w \left(\rd_\ga q_{fc} + 2\eps_{fci}\tl\rho^i\right) \w \left(\rd_\ga q_{ge} + 2\eps_{gel}\tl\rho^l\right),
\neqa
where we used $\eps_{abd} \hat q^{bf} \hat q^{dg} = \eps^{hfg}q_{ah}$. 
Of these four terms, consider now the two containing $\eps_{gel}\tl\rho^l$. We have 
\equ
\eps_{gel}\eps^{hfg}q_{ah}\hat q^{ce} \Si^a_\eps \w \tl \rho_{fc} = 
(\d^c_a \d^f_l - q_{al} \hat q^{cf}) \Si^a_\eps \w \tl \rho_{fc} \equiv 
q_{lb} \rho^b{}_a \w \Si^a_\eps - q_{al} \hat q^{cf} \Si^a_\eps \w \tl\rho_{fc} \equiv 0,
\nequ
where the first term vanishes thanks to  \Ref{Glaw2}, and second to the unimodularity of $q_{ab}$,
\equ\label{ddet0}
\hat q^{cf}  \tl\rho{}_{fc} = \f12 \hat q^{fc} \rd_{\ga}q_{fc} \equiv \f12 \rd_{\ga} ({\rm det} q_{ab}) = 0.
\nequ
We are left with 
\eqa\label{paperoga}
&& -\f\eta8 \eps^{hfg}q_{ah}\hat q^{ce} \Si^a_\eps \w \left(\rd_{\ga} q_{fc} + 2\eps_{fci}\tl \rho^i\right) \w \rd_{\ga}q_{ge}=
-\f\eta{16} \eps^{hfg} q_{ah} \hat q^{ce} \eps^{\mu\nu\rho\sigma} \Si_\eps^g{}_{\rho\sigma} D_\mu q_{fc} D_\nu q_{ge} + \nn \\ &&\qquad 
-\f\eta4 (q_{ai} \hat q^{ge} - \d^e_a \d^g_i) \Si^a_\eps \w \tl \rho^i \w \rd_{\ga}q_{ge}.
\neqa
The first term in the square bracket vanishes again as for \Ref{ddet0}. The second term gives
\eqa\label{paperino1}
\f\eta4 \Si^a_\eps \w \tl \rho^b \w \rd_{\ga}q_{ab} &=& \f\eta{16} \eps \eps^{\mu\nu\rho\sigma} \Si_\eps^a{}_{\mu\nu}
\left[ \d^{b(c}\Si^{d)}_\eps{}_{\rho\la}-\f12\d^{cd}\Si^{b}_\eps{}_{\rho\la} \right]\n^\la q_{cd}\n_\sigma q_{ab} = \nn\\
&=& \f{\eta}8 e \n^\sigma q_{ab} \n^\la q_{cd} \left[g_{\sigma\la}\left(\d^{ad}\d^{bc}-\f12\d^{ab}\d^{cd}\right) 
+\eps \d^{bc}\eps^{ad}{}_r \Si_{\eps}^r{}_{\sigma\la} \right],
\neqa
where in the last step we used \Ref{double}. Inserting \Ref{paperino1} into \Ref{paperoga} and massaging the indices we obtain $\f\eta8 e \n^\mu q_{ab} \n^\nu q_{cd} C^{abcd}_\eps{}_{\mu\nu}$ with $C^{abcd}_\eps{}_{\mu\nu}$ defined as in \Ref{C}. This completes the derivation of \Ref{SF} in the main text.


\end{document}